      [1999/12/01 v1.4c Il Nuovo Cimento]
\documentclass[varenna]{cimento}

\usepackage{graphicx}  
\title{Critical Overview of Agent-Based Models for Economics}
\author{M.~Cristelli, L.~Pietronero, \atque A.~Zaccaria }

\institute{ISC-CNR\\ Via dei Taurini 19, 00185, Roma, Italy.}
\institute{Dipartimento di Fisica\\ Sapienza, Universit\`a di Roma\\ P.le Aldo Moro 2, 00185, Roma, Italy.}

\begin{document}

\maketitle
\begin{abstract}
We present an overview of some representative Agent-Based Models in Economics. We discuss why and how agent-based models represent an important step in order to explain the dynamics and the statistical properties of financial markets beyond the Classical Theory of Economics. We perform a schematic analysis of several models with respect to some specific key categories such as agents' strategies, price evolution, number of agents, etc. In the conclusive part of this review we address some open questions and future perspectives and highlight the conceptual importance of some usually neglected  topics, such as non-stationarity and the self-organization of financial markets.
\end{abstract}
\section{Introduction}
We present a critical discussion of some of the most representative Agent-Based Models (ABM hereafter) of Economics. We embed them in a framework which highlights their key principles and the relative strengths and weaknesses. The objective is to point out the possible lines of convergency and improvement in order to focus towards an optimal model. A crucial weakness in this respect is that the experimental framework defined by the so-called Stylized Facts is still rather limited and an improvement of this body of knowledge appears to be the bottle neck in the field.
\\
\subsection*{ \textbf{Classical Theory of Economics}}
\begin{figure}[h!]
\begin{center}
\includegraphics[scale=0.45,angle=0]{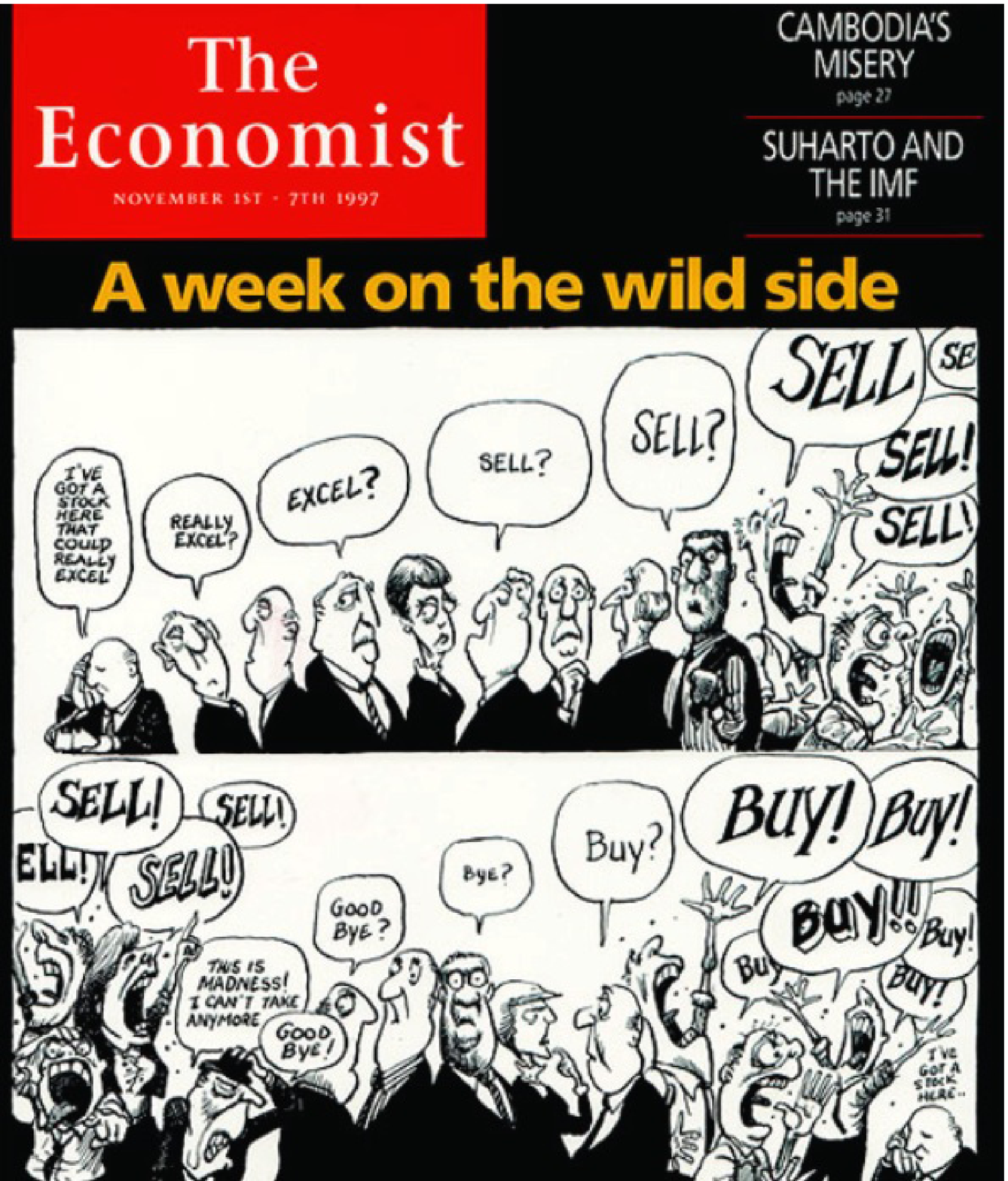}     
\caption{\label{eco_cart}This cartoon appeared on the cover of The Economist on November 7, 1997 to illustrate non rationality, limited information and a strong herding behavior. All these elements are absent in the standard model of Economics. The ambition of the agent-based models is to include this type of elements in the modeling of Economics.}
\end{center}
\end{figure}

The Classical theory of Economics is based on the following elements \cite{buchanan,kreps}
\begin{itemize}
\item Situation of equilibrium with equal (representative) agents which are (quasi) rational, have the same information and process it in the same way;
\item Important price changes correspond to new information which arrives on the market. The fact that this information is random and independent leads to the famous random walk model \cite{bachelier,feller} and the corresponding Black and Scholes equations \cite{blacknscholes,hull};
 \item This new information modifies the ratio between offer and demand and then also the price. This corresponds to a mechanical equilibrium of the market;
\item These concepts also imply a Cause-Effect relation in which large price changes are due to the market reaction to the arrival of exogenous important news. Therefore a large price change is supposed to be associated to an equally large exogenous event.
\end{itemize}
It is too easy to argue that most of these assumptions have no basis at all. So one may wonder why they are so widely adopted. The real reason is that they permit an analytical treatment of the problem. This is a very different perspective from natural sciences in which very few realistic problems can be treated exactly but still one can get an appreciable level of understanding with suitable approximation schemes or computer simulations. In addition these ideas are usually not tested against empirical data; a fact that strongly limits the scientific basis of the whole framework.\\\\
The most reasonable of the above assumptions is the fact that indeed many external news are random and incoherent so the random walk appears reasonable as a simple modeling. On the other hand the assumptions made for the behavior of the agents are very far from reality. A famous cartoon of the Economist to stress this point is reproduced in fig. \ref{eco_cart}. The agents can be very different from each other, their level of information and the way the use it is also very different. Finally they are not at all independent and, especially in situations of crisis the rationality hypothesis can be seriously wrong. In these situations fear and panic, as well as euphoria, lead to very strong herding behavior (fig. \ref{eco_cart}) which is completely neglected in the standard model.
\\\\
From an empirical point of view one can observe marked deviations from the standard picture:
\begin{itemize}
\item great catastrophic events like the '87 crash, the Internet bubble of 2000 and the recent case of the subprimes do not seem to have any relation with specific events or new information;
\item there are clear deviations from the gaussian behavior for the price distribution which are named Stylized Facts (SF). These can be observed for the large price fluctuations but are actually present at all scales, even in relatively calm periods;
\item breaking of the Cause-Effect relation. Often large price fluctuations are not associated to any special event. In table \ref{tab_price} we report the top ten price movements from the second world war til 1989 together with their origin as reported by the media. Even for the recent subprime crisis the total amount of money corresponding to the subprimes is very small with respect to the catastrophyc effect it had worldwide.
\end{itemize}
\begin{table}
  \caption{We report the ten largest price movement in the Standard \& Poor's index from the second world war till 1989 together with the official explanation of their ``origin'' as reported in the media (see \cite{cutler} for further details). }
  \label{tab_price}
  \begin{tabular}{ccl}
\hline
Date      & Percent Change &New York Times Explanation  \\
\hline
\hline
Oct. 19, 1987        &-20.47    & Worry over dollar decline and trade deficit \\
Oct. 21, 1987          &+9.10    & Interest rates continue to fall, bargain hunting  \\
Oct. 26, 1987   &-8.28 & Fear of budget deficit, margin calls \\
Sep. 3, 1946   &-6.73 & ``... No basic reason for the assault on prices'' \\
May 28, 1962   &-6.68 &  Kennedy forces rollback of steel price hike\\
Sep. 26, 1955   &-6.62 & Eisenhower suffers heart attack \\
Jun. 26, 1950   &-5.38 & Outbreak of Korean War \\
Oct. 20, 1987   &+5.33 & Investors looking for ``quality stocks'' \\
Sep. 9, 1946   &-5.24 & Labor unrest in maritime and tracking industries \\
Oct. 16, 1987   &-5.16 & Fear of trade deficit and tension with Iran. \\
    \hline
  \end{tabular}
\end{table}
\subsection*{ \textbf{Towards Complexity}}
This situation naturally calls for a possible description in terms of critical phenomena and complex systems \cite{baffoEN}. The study of complex systems refers to the emergence of collective properties in systems with a large number of parts in
interaction among them. These elements can be atoms or macromolecules in a physical or biological context, but also people, machines or companies in a socio-economic context. The science of complexity tries to discover the nature of the
emerging behavior of complex systems, often invisible to the traditional approach, by focusing on the structure of the interconnections and the general architecture of systems, rather than on the individual components.
This implies a change of paradigm form the previous mechanical model to a complex model in which intrinsical instabilities can develop in a self-organized way without a cause-effect relation \cite{baffoEN,buchanan, bouchaudnature, farmernature}.
\\
\subsection*{ \textbf{Why Agent-Based Models}}
The so-called ABM represent a broad class of models which have been introduced to describe the economic dynamics in a more realistic way. Their building blocks are:
\begin{itemize}
\item the agents are heterogeneous with respect to their various properties like strategies, wealth, time scale, etc;
\item the interaction between them is a fundamental element and, of course, it can have many different characteristics;
\item price dynamics depends on the balance between offer and demand but the specific implementation can be different.
\end{itemize}
Usually the dynamics of this type of models is able to produce deviations from the gaussian behaviour related to the SF for some specific range of parameters.\\
Additional elements which are often neglected but are important in our opinion are the fact that the total number of agents or strategies can strongly vary in time and that the system has to evolve spontaneously towards the quasi-critical state which shows the SF.\\\\
\begin{figure}
\begin{center}
\includegraphics[scale=0.45,angle=90]{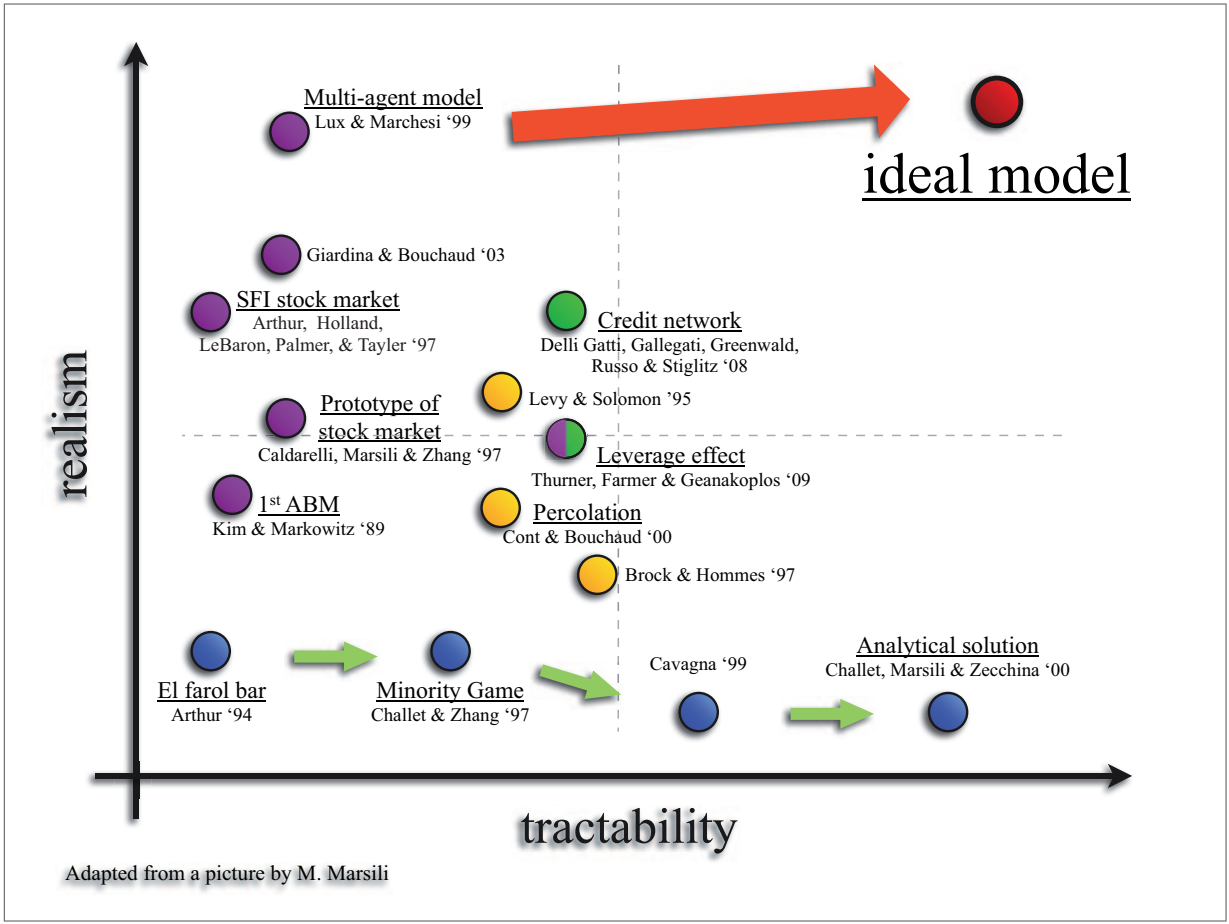}     
\caption{\label{revstr}In this picture we report several ABM organized along the two axis of  realism and tractability. The picture was inspired by a talk by Matteo Marsili and is intended to show the rather different perspectives of the various models together with the path we have tried to follow in constructing our approach. The Ideal Model is still far and the available data (Stylized Facts) are too scarce to permit a quality assessment of the various models.}
\end{center}
\end{figure}
The paper is organized as follows.\\
In \textbf{section 2} we report the main statistical properties of financial markets.
\\In \textbf{section 3} we briefly describe the classical economical theory focusing our attention on the consequence of this assumptions applied to financial markets.\\
In \textbf{section 4} we analyze some ABM with respect to specific features such as the type of strategies, the number of agents, the self-organization, etc.\\
In \textbf{section 5} we discuss a different approach to ABM in Economics presenting a recently introduced model which is the first step towards a systematic analysis of financial market features with the final aim of defining the characteristic properties of stock markets from the models, and then to identify their role and importance in real financial markets.\\
In \textbf{section 6} we give some perspectives of development of agent-based modeling.
\section{\textbf{Stylized Facts}}
The name SF refers to all non trivial statistical evidences which are observed throughout financial markets. Almost all price time series of financial stocks and indexes approximatively  exhibit the same statistical properties (at least qualitatively). In addition it has been shown that SF are robust on different timescales and in different stock markets \cite{rama}. \\
The systematic study of SF has begun in very recent time (approximately from '90) for two reasons: a technical and a cultural one. The former one  is that the huge amount of empirical data produced by financial markets are now easily available in electronic format and can be massively studied thanks to the growth of computational power in the last two decades.
In order to make a comparison with some traditional fields of Physics, a similar quantity of information is observed only in the output of a big particle accelerator.
The latter instead is due to the fact that traditional approaches to economical systems neglect empirical data as candidates respect to which a theory must be compared differently from Physics. From this point of view standard Economics is not an observational science.
\\\\
Turning now our attention to the experimental evidences of financial markets, the main SF are
\begin{itemize}
\item the absence of simple arbitrage,
\item the power law decay of the tails of the return distribution,
\item the volatility clustering.
 \end{itemize}
 In the following sections we analyze them.
\subsection{\textbf{Absence of simple arbitrage}}
\begin{figure}[h!]
\begin{center}
\includegraphics[scale=0.45]{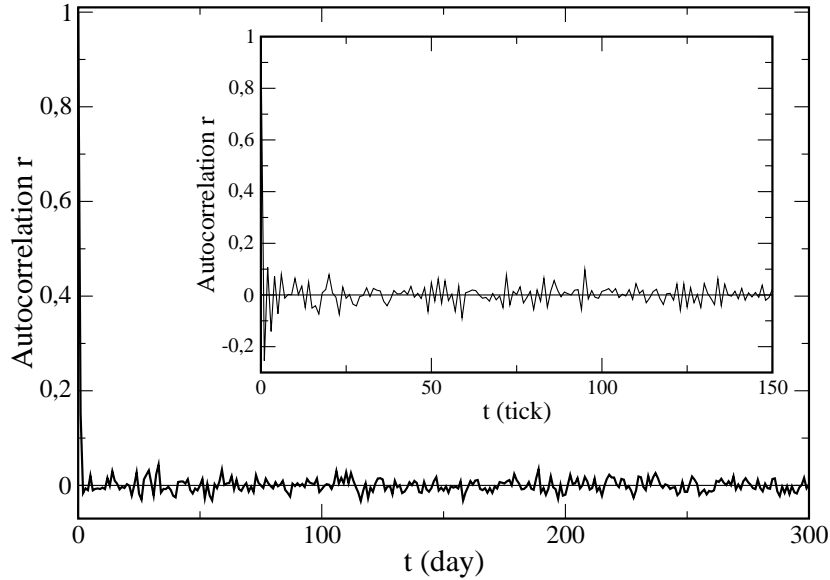} 
\caption{ \label{fig_autocorrel} We report the autocorrelation function of returns for two time series. The series of the main plot is the return series of a stock of New York Stock Exchange (NYSE) from 1966 to 1998 while the series of the inset is the return series of a day of trading of a stock of London Stock Exchange (LSE). As we can see the sign of prices are unpredictable that is the correlation of returns is zero everywhere. The time unit of the inset is the tick, this means that we are studying the time series in event time and not in physical time. }
\end{center}
\end{figure}

The absence of simple arbitrage in financial markets means that, given the price time series up to now, \textbf{the sign of the next price variation is unpredictable on average}. In other words it is impossible to make profit without dealing with a risky investment. This implies that the market can be seen as an open system which continuously reacts to the interaction with the world (i.e. trading activity, flux of information, etc) and self-organizes in order to quickly eliminate arbitrage opportunities. This property is also called \textbf{arbitrage efficiency}.\\\\
This condition is usually equivalent to the informational efficiency expressed in Economical literature saying that the process described by the price $p_t$ is a martingale that is 
\begin{equation}
E[p_t|p_s]=p_s
\end{equation}
where $t>s$. Here we are assuming that the price is a synthetic variable which reflects all the information available at time $t$. If this is not true the conditioning quantity is the available information $I_s$ at time $s$ and not only the price $p_s$. \\\\
However, the condition of martingale is uneasy from a practical point of view and \textbf{the two-point autocorrelation function of returns is usually assumed as a  good measure of the market efficiency}
\begin{equation}\label{autocorr}
\rho(\tau,t)=\frac{E[r_tr_{t+\tau}]-E[r_t]E[r_{t+\tau}]}{E[r_t^2]-E[r_t]^2}.
\end{equation}
If the process $\{r_t\}$ is at least weakly stationary then eq. \ref{autocorr} simply becomes $\rho(\tau)=(E[r_tr_{t+\tau}]-\mu_r^2)/\sigma_r^2$ where $\mu_r=E[r_t]$ and $\sigma_r=E[r_t^2]-E[r_t]^2$. If the autocorrelation function of returns is always zero we can conclude that the market is efficient. \\
\textbf{In real markets the autocorrelation function is indeed always zero} (see fig. \ref{fig_autocorrel}) except for very short times (from few seconds to some minutes) where the correlation is negative (see inset of fig. \ref{fig_autocorrel}). The origin of this small anti-correlation is well-known and due to the so-called \textit{bid ask bounce}. This is a technical reason deriving from the double auction system which rules the order book dynamics (see \cite{libroandrea} for further details). \\\\
In the end we want to stress that the efficiency is a property that holds on average: locally some arbitrage opportunities can appear but, as they have been exploited, the efficiency is restored \cite{taka2,taka1}. 
\subsection{\textbf{Fat-tailed distribution of returns}}
\begin{figure}
\begin{center}
\includegraphics[scale=0.45]{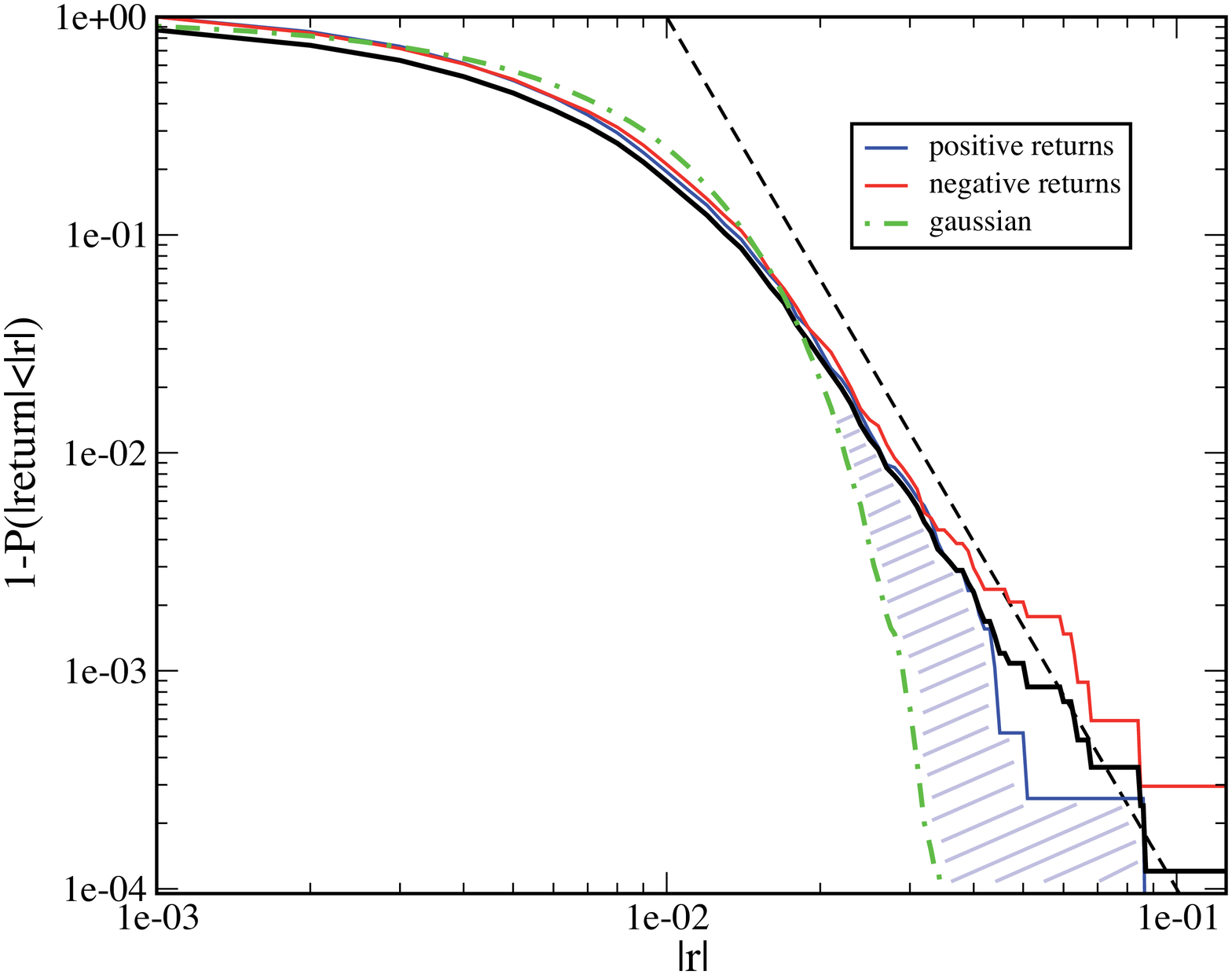} 
\caption{\label{fattails}We report the complementary cumulative distribution function of the absolute value of returns (solid black line). The green dashed line ($\cdot -$) is the complementary cumulative distribution function of a gaussian with the same variance of the real return distribution. The dashed black line is a pure power law decay with exponent $\alpha=4$. The blue and red lines are instead the complementary cumulative distribution functions for positive and negative returns respectively. We can see that red curve has a slower decay with respect to the blue one. This asymmetry between positive and negative returns is the origin of the non zero skewness of the probability density function of returns. }
\end{center}
\end{figure}

\textbf{The distribution of price variations} (called returns) \textbf{is not a Gaussian} and prices do not follow a simple random walk. In details very large fluctuations are much more likely in stock market with respect to a random walk and dramatic crashes are approximately observed every $5-10$ years on average. These large events cannot be explained by gaussian returns. Therefore to characterize the probability of these events we introduce 
the complementary cumulative distribution function $F(x)$
\begin{equation}
F(x)=1-\textrm{Prob}(X<x)
\end{equation}
which describes the tail behavior of the distribution $P(x)$ of returns.
\\\textbf{The complementary cumulative distribution function $F(x)$ of real returns is found to be approximately a power law} $F(x)\sim x^{-\alpha}$ with exponent in the range $2-4$ \cite{bouchaud}, i.e. the tails of the probability density function (pdf) decay with an exponent $\alpha+1$. \textbf{Since the decay is much slower than a gaussian this evidence is called Fat or Heavy Tails}. Sometimes a distribution with power law tails is called a Pareto distribution. The right tail  (positive returns) is usually characterized by a different exponent with respect to the left tail (negative returns). This implies that the distribution is asymmetric in respect of the mean that is the left tail is heavier than the right one ($\alpha^+>\alpha^-$).\\
Moreover the return pdf is a function characterized by positive excess kurtosis, a Gaussian being characterized by zero excess kurtosis. 
In fig. \ref{fattails} we report the complementary cumulative distribution function $F(x)$ of real returns compared with a pure power law decay with exponent $\alpha=4$ and with a gaussian with the same variance. \\\\
When the tail behavior of the return distribution is studied varying the time lag at which returns are performed  \cite{rama}, a transition to a gaussian shape is observed for yearly returns. However it is unclear if this transition  is genuine or due to a lack of statistics or to the non stationary return time series.
\subsection{\textbf{Volatility clustering}}
\begin{figure}
\begin{center}
\includegraphics[scale=0.45]{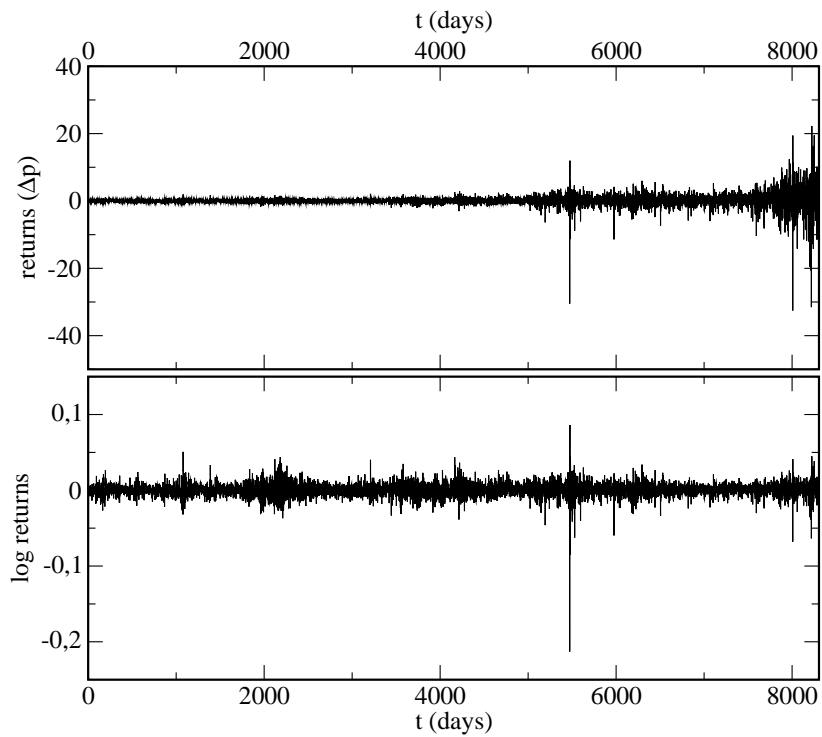} 
\caption{\label{vol_clust1}Return time series of a stock of NYSE from 1966 to 1998. The two figures represent the same price pattern but returns are differently computed. In the top figure returns are calculated as simple difference, i.e. $r_t=p_t-p_{t-\Delta t}$ while in the bottom one returns are log returns that is $r_t=\log p_t - \log p_{t-\Delta t}$. From the lower plot we can see that volatility appears to be clustered and therefore large fluctuations tend to be followed by large ones and vice versa. The visual impression that the return time series appears to be stationary for log returns suggests the idea that real prices follow a multiplicative stochastic process rather than a linear process.}
\end{center}
\end{figure}

\begin{figure}
\begin{center}
\includegraphics[scale=0.45]{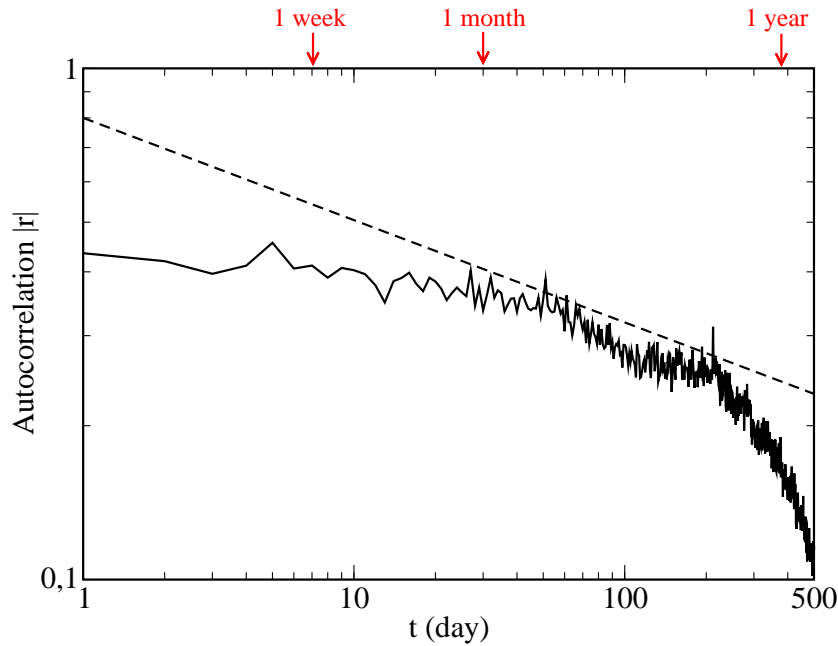}    
\caption{\label{vol_clu_fig}Autocorrelation function of volatility measured as the absolute value of returns. We find that the function can be approximately described by a power law. We report a pure power law decay with exponent $0.2$ for comparison. The return time series used in this analysis is the previous NYSE one from 1966 to 1988. It is worth noticing that the volatility is significantly correlated and then clustered on time scale longer than one year.}
\end{center}
\end{figure}

In the lower panel of fig. \ref{vol_clust1} we report the return time series of a NYSE stock (returns are here defined as $\log (p_{t+1}/p_t)$). As we can see \textbf{the behavior of returns appears to be intermittent} in the sense that periods of large fluctuations tend to be followed by large fluctuations regardless of the sign and the same behavior happens for small ones. 
\\In Economics the magnitude of price fluctuations is usually called volatility. It is worth noticing that a clustered volatility does not deny the fact that returns are uncorrelated (i.e. arbitrage efficiency). In fact correlation does not imply probabilistic independence, while the contrary is true. Therefore the magnitude of the next price fluctuations is correlated with the present one while the sign is still unpredictable. In other words \textbf{stock prices define a stochastic process where the increments are uncorrelated but not independent}.\\\\
Different proxies for the volatility can be adopted: widespread measures are the absolute value and the square of returns. As a consequence of the previous considerations about the clustering of volatility, \textbf{the autocorrelation function of absolute  (or square) returns is non zero}. We also find that \textbf{the autocorrelation is well-described by a power law decay} with exponent ranging from $-1$ to 0 as reported in in fig. \ref{vol_clu_fig}. The very slow decay means that volatility is correlated on very long time scales from minutes to several months/years. The exponent of the autocorrelation function is not universal as the one of fat tails but it is typically around $0.2 - 0.3$. The volatility clustering was observed the first time by Mandelbrot in 1963 \cite{benoit1}.
\subsection{\textbf{Other SF}}
Beyond these SF we can state other relevant effects which are widespread in financial markets such as 
\begin{itemize}
\item the gain/loss asymmetry, i.e. one observes large drawdowns in stock prices and stock index values but not equally large
upward movements. This is linked to the asymmetry of the return pdf.
\item leverage effect: the volatility of an asset
are negatively correlated with the returns of that asset.
\item
trading volume and volatility are correlated.
\end{itemize}
See also \cite{rama,taka2,taka1,bouchaud,mantegnabook,reviu_abergel} for more details about SF and their analysis.
\subsection{\textbf{Stationarity and time-scales}}
Before turning our attention to the analysis of the models which try to interpret SF, we want to discuss a final question: the stationarity and the time scales of the observation of financial markets.\\
The hypothesis of stationarity is usually invoked because, if satisfied,  the statistical properties of the phenomenon under consideration become invariant under temporal translation. In the case of financial markets it is not clear whether the return time series verifies this condition: \textbf{intraday activities, seasonality, weekends, holidays, Economy growth are elements that can} \textit{a priori} \textbf{make the returns not stationary}. However, it can be argued that the process is observed on the wrong scale (see \cite{lebraon_stat}) and on a larger time scale the process may become stationary.\\\\
Last but not least it has been hypothesized that the non stationarity of financial series derives from the fact that they are studied in physical time units. On this account it has been proposed to define a time rescaling such that the transformation which makes stationary the financial data is the correct one. However, the choice of which elements should be involved in this transformation is arbitrary and ranges form seasonality of the calendar to the volumes of trading activity. The question is still open (see \cite{rama_ref1,rama_ref2,rama_ref3,rama_ref4} for further references).
\section{\textbf{Standard Economic Theory}}\label{SET}
In the previous section we have reported the main empirical features of financial markets. Now we briefly review the so-called neoclassical framework which has been the dominant paradigm of economical theory for fifty years.\\
\textbf{According to standard economical theories markets are in an equilibrium state}. \\\\
Before analyzing the consequence of such an assumption, let us briefly describe how an economy is modeled by the neoclassical approach. 
\textbf{The economical system is composed of goods, agents and firms}. Agents and firms own a certain amount of goods, have a set of technologies to transform goods in other goods and an utility function quantifies agents and firms' preferences in good consumption. Agents and firms only aim at maximizing their utility by selling, buying, producing or consuming goods. \\\\
The typical hypothesis of equilibrium models are: 
\begin{enumerate}
\item a single \textbf{agent's action is marginal} with respect to the price, that is the price variations due to the single agent is a contribution of higher order with respect to the price variation;
\item agent's aim is the \textbf{maximization of the utility} given the price vector of goods;
\item the price vector is the solution of the equation that makes the aggregate demand and the aggregate supply equal; 
\item \textbf{agents are fully rational} and are fully informed, that is they have perfect models of reality to determine their actions and know the price vector.
\end{enumerate}
There are two main motivations of the great success of equilibrium models.\\
The former is that equilibrium models are a closed and relatively simple quantitative theory for economical system. The models are closed in the sense that given the previous four assumptions no extra ingredients are needed to determine the equation connecting goods, technology, production, trade and prices.\\
The second reason is that if an extra hypothesis is added to the behavior of the utility function and of technology, then the equilibrium state always exists \cite{arrow}. The demonstration of the existence of this equilibrium state consists in the demonstration of the existence of a fixed point of a suitable map acting on the space of price vectors. The two additional hypothesis can be summarized saying that the more an agent consumes a good, the less the agent will obtain a benefit from the consumption of an extra unit of that good and that the more a technology is used to produce a good, the less the extra production derives from one extra unit of input. We do not report the problems that arise from the question of how this fixed point is reached. We only highlight that the existence of this point does not guarantee that the system is always able to reach this state. \\\\
The models herein proposed are essentially one-time models. In their original formulation they are static and do not deal with the dynamics of the price evolution and of agent's allocation of goods. However, it has been shown that equilibrium models can be extended to a multi-time set up and even to an economy composed of financial goods and not only of concrete and physical goods. The details of these two features go beyond the goal of the present review. The readers who are interested in these questions can find more information in \cite{kreps,ref_far1,ref_far2}. \\\\
Now we want to analyze the consequence of the equilibrium assumptions in a stock market. Because of the multi-time horizon we have to introduce a new hypothesis and we are going to see that it is the stronger assumption made until this point. In a multi-time set up, \textbf{agent's rationality implies the knowledge of all the present prices but also of all the dependences of prices on the possible future states on which the agents are uncertain} (these states are called states of Nature). Furthermore the extension from a one-time to a multi-time horizon implies that \textbf{the maximization of the utility function must now depends on all the possible future scenarios}. Thus, whether or not a good is consumed today depends on all the possible future history of price vector and on all the probabilities of the uncertain states of Nature. Thus the consequence of the full rationality become a bit strange with respect to our daily experience about preferences in uncertain conditions. In particular in equilibrium models we have to assume that the probability of the uncertain states of Nature are known or perfectly deduced by rational agents at the beginning of the history. This also implies that the uncertain quantities are stationary in time and on all time scales. \\
On the other end it is simple to show that in the original one-time equilibrium models the equilibrium state is also a Pareto efficient state. From this point of view the model is allocative efficient. In fact a state is defined Pareto efficient if there is not any change of the allocation of all agents that make at least one agent happier with respect to his/her utility. The extension to financial markets is far from being trivial because the equilibrium state in a multi-time horizon and in uncertain conditions is not usually a Pareto efficient state.
Therefore rationality can be extended to financial contest even if it includes the unrealistic perfect knowledge of all future sources of uncertainty while efficiency is in general not satisfied by equilibrium model of financial markets.  \\\\
For this reason a weaker efficiency is usually required to hold in financial contest: the informational efficiency which we have introduced in the previous section. In mathematical terms the informational efficiency means that the price process is a martingale with respect to the information process $I_t$. In its stronger version the price is a synthetic variable which incorporates all the information and therefore present price is the best prediction of any future price. \\The most striking consequence of \textbf{the full rationality assumption and of perfect informational efficiency} is that \textbf{prices should follow the fundamental value} of the financial good that is the future discounted forecast of the dividend payed by the stock. In this market price and fundamental price would always coincide (see the brief argument in \cite{sf2} to illustrate this point). We want to stress that all agents are fully rational and endowed with the same information, the logical consequence is that all agents, being rational, get the same forecast about future. Thus agents are rational and homogeneous and this permit to deductively forecast the future prices of a stock. This procedure leads to the formula
\begin{equation}
p_t=\sum_{k=1}^\infty\beta^kE[d_{t+k}|I_t]
\end{equation}
where $\beta$ is the discount factor and is equal to $1/(1+r)$ where $r$ is the risk free rate (i.e. treasure bonds).\\
We can see that the only allowed fluctuations of prices derive from the uncertainty due to the conditioning of the present information $I_t$. So that we have found  prices will follow the fundamental value except for small deviations from it due to fluctuations of information. Moreover in such a context there are not any incentives to trade since there are not any speculative opportunities. The investors will gain profit only from the revenues of the dividends. 
\\\\
If we compare the predictions of standard neoclassical models with the observational results presented in the previous section \textbf{we find that SF can be hardly fitted by the framework of equilibrium theory}. Even worse, equilibrium theory does not appear as the natural starting point for a new theory. Instead \textbf{a paradigm shift} is required and ABM are the first step in this direction.  Actually, as equilibrium models, ABM are more similar to metaphors than to real predictive/quantitative models but their advantage relies in the fact that SF are their starting point and their benchmark to determine their fitness in describing financial markets.  
\section{\textbf{Agent-Based Models}}
In this section we present and discuss eight ABM which, according to the authors of the present review, well represent the development of this field. Differently from most reviews of this area \cite{reviu_abergel,reviu,reviu_hommes}, the analysis of each model is performed in a very schematic way according to some categories. These categories are:
\begin{itemize}
\item \textbf{Aim of the model:} we describe the reasons of the introduction of each model.
\item \textbf{Agents and strategies:} we give a detailed description of the type of agents involved in the model, the strategies used to trade, the mechanism which rules the change of strategy (in the model where this feature is present), etc.
\item \textbf{Number of agents:} we discuss if the number of active agents changes in time or is kept fixed.
\item \textbf{Price formation:} in this section we discuss the mechanisms through which price evolution and strategy evolution interacts
\item \textbf{Origin of Stylized Facts:} we analyze which elements give origin to SF.
\item \textbf{Discussion realism/tractability:} the development of an ABM sets a not trivial problem with respect to the 
compromise between tractability and realism. We review the advantages and disadvantages of each model in relation to this question.
\item\textbf{Self-Organized Criticality}: if it is possible, we analyze how the problem of the market self-organization in the regime of SF is addressed. 
\end{itemize}
\subsection{\textbf{Kim and Markowitz model: an attempt to explain the \textit{Black Monday}}}
\subsubsection{\textbf{Aim}} On Monday, October 19, 1987 stock markets worldwide underwent the largest drawdown of financial history in a very short time. In one day Dow Jones Industrial Average lost 508 points which correspond to a percentage drop of about $23\%$. The losses were even worst in the rest of the world, for instance in Hong-Kong, where the crash began, in United Kingdom, in New Zealand, etc. \\The origins of this crash are still unclear and a conclusive explanation is still missing \cite{black2,black1}. It is interesting to note a similar lack of conclusive answer is also observed for the recent Flash Crash on May 5, 2010. The report of the U.S. Securities and Exchange Commission (SEC) \cite{flashcrash} gives a detailed description of the event timeline but does not really answer how the Dow Jones could lose about $10\%$ in few minutes and, as fast as the drop, could recover approximately its initial value. \\\\
The Black Monday's causes immediately appeared and in some extent still appear rather \textit{mysterious} because \textbf{there was not any flux of information which could explain the origin of such a large and such a widespread negative fluctuation}. As we have already seen in sec. \ref{SET}, in the traditional framework   
a daily return which is dozens of times larger than the typical return simply should not exist. In fact according to the classical Economics picture of price dynamics the agents are fully rational and all endowed with the same information, markets are always liquid and by consequence they are substantially efficient and in equilibrium. If these assumptions were true then the price dynamics would trivially follow the fundamental value and the fluctuations of prices would simply reflect the gaussian fluctuations of the flux of incoming information and would be always and almost immediately rebalanced by rational market participants \cite{rational}. \\The Black Monday perhaps had the merit to reveal to the world \textbf{the inadequacy of the cause-effect relation between information and price fluctuations}. This inadequacy was already clear for some pioneering scientists at that time but the equilibrium and rationality assumptions had not been really questioned before this dramatic event.  In this respect 
we want to stress that the Black Monday losses are merely a consequence of one of SF that we previously mentioned, the fat tails that had been discovered about twenty years before by Mandelbrot \cite{benoit1}. In general almost all the SF hardly find a real explanation in the framework of traditional economical hypothesis. \\\\
\textbf{The research of the real causes beyond an anomalous flux of information was a perfect playground for the first modern multi ABM} proposed by Kim and Markowitz in 1989 \cite{KIM}. ABM usually do not have an underlying microeconomical foundation but they have turned out to be powerful tool in the last twenty years in order to go over the representative agent framework. ABM permit to easily investigate concepts such as heterogeneity, herding, illiquidity, lack of rationality. The firsts who really recognized the potential of ABM  were two economists, i.e. Kim and Markowitz, who proposed the model which we are analyzing in this section.\\
\subsubsection{\textbf{Agents and Strategies}}
The model is very focused on the causes of the Black Monday. In particular the authors of this model want to investigate whether the presence of hedging and portfolio insurance strategies \cite{portfolio_en} can destabilize the market or at least produce a significant increase of volatility. In fact at the very end the model tries to understand the link among volatility and portfolio insurance strategies. \\\\
In practice the model has two classes of agents. The former type is called \textit{rebalancer}, the latter \textit{portfolio insurer} and both classes can trade two assets. As in many ABM these two assets are a risk free asset and a risky one of which the return can be either positive or negative. In this model the risk free asset has a zero interest rate and we can identify it with cash. This choice for the interest rate produce a detrended price behavior. In fact the interest rate of the risk free bond introduces a deterministic trend in the price evolution which mimics the growth of global economy. Therefore there is not any loss of generality in setting this rate to zero with respect to the aim of the model. \\
Each agent is endowed with a finite wealth $w_t=q_tp_t+c_t$ where the set $(q_t,c_t)$ describes its portfolio composition at each time $t$: $c_t$ is the amount of capital invested in cash (the risk free bond) while $q_t$ is the number of stocks held and $p_t$ is the price of the stock. \\
The composition of the portfolio $(q,c)$ depends on the agent's strategy. \\\\
The first class operates in order to keep the proportion invested in cash and in the risky stock constant given the constraint of finite wealth $w_t=q_tp_t+c_t$:
\begin{equation}\label{rebal}
\frac{q_tp_t}{c_t}=\alpha.
\end{equation}
In the model $\alpha$ is set equal to $1$ so that rebalancers invest half of their capital $w_t$ in cash and half in stocks. \\\\
The rule to define the portfolio composition of portfolio insurers is more sophisticated. This kind of agents aims at keeping constant the following proportion, given the constraint deriving from finite wealth, 
 \begin{equation}
\frac{q_tp_t}{w_t-f}=\beta
\end{equation}
where $f$ is a fraction $\gamma$ of the initial capital $w_0$ and $\beta>1$. This time the proportion is performed with respect to the wealth of the agents decreased by the minimum amount $f$ of the capital which these agents want to have guaranteed. \\\\
\textbf{The rebalancers have a stabilizing effect} on the price behavior because if price goes up then $q_tp_t$ also grows, therefore in order to verify the condition given by eq. \ref{rebal} they update their portfolio $(q,c)$ reducing their exposition in the risky asset. In sec. \ref{feedKM} we are going to see that the mechanism for the price evolution assumes that the price variation is produced by the supply/demand imbalance. In this way the price will tend to decrease at the following step. Obviously the opposite update strategy of $(q,c)$ holds if the price goes down. \\
Instead \textbf{portfolio insurers can destabilize the market} because their response to a price variation is the opposite with respect to rebalancer action. In fact if the price of the risky stock grows, their wealth grows too. Then the update of the portfolio consists in a higher exposure in the risky asset and consequently their action will continue to fuel the price growth and viceversa if price decreases. The choice of a constant $\beta$ larger than $1$ amplifies the response of this type of agents when they update their portfolio composition.\\
In sec. \ref{feedKM} we describe in details how the portfolio $(q,p)$ of each agent is reviewed.\\\\
The model is characterized by heterogeneous strategies (i.e. two ones) but there is not a real heterogeneity inside the two types of agents because all rebalancers and all portfolio insurers respectively share the same parameters for their strategies, ($\alpha$ for rebalancer, $\beta,\gamma$ for insurers).\\\\
Changes of strategies are not allowed in this model. The unique change consists in removing those agents whose wealth is zero. 
\subsubsection{\textbf{Number of agents}}
The number of agents is trivially dependent on the time because if the wealth of an agent goes to zero then she is removed from the market. We recall that the wealth value at the end of each trading day  is $w_t=q_tp_t+c_t$ where $p_t$ is the closing price of that day. Consequently the number of players is always non increasing. We said that this dependence is trivial because this model does not aim at investigating this feature and therefore it substantially belongs to the class of ABM characterized by a fixed number of agents.
\subsubsection{\textbf{Price formation}}\label{feedKM}
The price evolves according to the imbalance between supply and demand due to the updates of the portfolio compositions.  The mechanism for the  execution of orders resembles a very simple order book-like mechanism which permits to take into account the effect of finite liquidity. Orders due to the review of the portfolio composition are executed only when a buying request is matched by a selling request at the same price. 
\\\\Moving to the details of such a mechanism, an agent decides to update her/his portfolio at random interval. This means that the order submission is asynchronous as in real dynamics of order books and trades take place at discrete time. In order to decide how to rebalance the portfolio the investors must estimate the next price and this forecast is called $p_{est}$. We want to stress that even if all agents of a class share the same parameters, the update of their portfolio is in general different because of the asynchronous update rule and consequently because of the different configurations of orders encountered to estimate the next price. 
Four scenarios can happen:
\begin{itemize}
\item There are not orders at this time: $p_{est}$ is assumed equal to the previous trading price
\item There are only buy orders stored in the market: the investor who is currently reviewing his/her portfolio expects a price increase and $p_{est}=1.01 \max(p_{buy} )$ where $\max(p_{buy})$ is the highest buy order stored.
\item There are only sell orders: the investor who is currently reviewing his/her portfolio expects a price decrease and $p_{est}=0.99 \min(p_{sell} )$
\item There are both sell and buy orders: the investor set $p_{est}=[\max(p_{buy})+\min (p_{sell})]/2$
\end{itemize}
The amount of stocks to sell or buy in order to rebalance the portfolio is fixed by the difference between $q_tp_{est}$ and the target value set by the strategy of the agents. For instance if the agent is a rebalancer and $q_tp_{est}>w_t$ the agent will put a sell order, viceversa if the opposite inequality holds. The price of the submitted order (both for insurers and  rebalancers) is equal to $0.99p_{est}$ if the agent is selling, to $1.01p_{est}$ if the agent is buying. 
The order can be immediately matched if an order of the opposite type is present in the market and the order execution fixes the new trade price, otherwise the order is stored. \\
A trading day is declared closed when every investors has reviewed their portfolio at least one time and the agents whose wealth $w_t$ is equal to zero at the closing price are removed from the simulation. 
In \cite{reviu} it has been shown that this rather complicated mechanism for the price update can be replaced by a more easier relationship among price variation and excess demand and the authors of \cite{reviu} find approximately the same results. 
\subsubsection{\textbf{Origin of SF}} This model does not directly deal with the problem of SF as we have already mentioned. Some later studies (for instance see \cite{reviu}) check whether the model is able to reproduce some SF. The conclusion is negative but we cannot say that this is really a negative point because we are testing the model on a task for which it is not designed.\\\\
Let us move to a brief review of the main results of this model. \textbf{The simulated market exhibits increasing volatility and trading volume for increasing values of the fraction of portfolio insurers}. The main drawbacks of the model are the quasi-periodic behavior of prices when the fraction of portfolio insurers is large. In \cite{reviu} it has been shown that there exists a positive correlation between the fraction of bankruptcy and the initial fraction of portfolio insurers and between the fraction of bankrupt insurers and the fraction of initial portfolio insurers.  They also show that if the number of initial agents is increased then the fraction of bankruptcy is reduced and consequently the volatility grows with respect to a case with less agents. \\
In the end a trivial decreasing volatility in time is observed because of the removing rule of zero-wealth agents. 
\subsubsection{\textbf{Discussion realism/tractability}} The model was introduced to study the stability of financial markets with respect to agent strategies. Thus, being the first real ABM, the model has the merit for searching endogenous causes of the instabilities of markets. From this point of view this work is pioneering and opens a novel vision and approach of market problem. Nevertheless the absence of replacements for the bankrupt agents and the quasi-periodic pattern of prices reduces its success in explaining market instabilities. \\\\
Furthermore the model makes a set of too simplistic assumptions  and introduces an unnecessary degree of complication for the mechanism of price evolution. In \cite{reviu} the authors show that similar results can be obtained with a much easier price rule. Therefore the complication of an order book-like mechanism for the trading activity does not introduce any real \textit{added-value} and only reduces the tractability of the model. But we want to stress that liquidity  is not only a cosmetic ingredient in the framework of multi-agent modeling. We only argue that in the this particular context the complicated mechanism for the order matching does not have any concrete effect on the main task the model was designed for. This aspect shows \textbf{a typical problem of agent-based modeling: the optimal trade-off between realism and essential ingredients} in order to understand the matter in hand. In other models instead we will highlight the trade-off between tractability and realism (see Fig. \ref{revstr}). 
\subsubsection{\textbf{Self-Organized Criticality}}
This aspect is not addressed in this model.
\subsection{\textbf{Santa Fe Artificial Stock Market} }
\subsubsection{\textbf{Aim}} The Santa Fe artificial stock market  is the result of the collaboration of two computer scientists, a physicists and two economists at Santa Fe Institute \cite{sf2}. In this section we will analyze the model presented in \cite{sf2} which is the most interesting version with respect to SF and complexity of financial markets. However, those interested in an earlier report about Santa Fe artificial stock market can see \cite{sf1}. The work proposed in \cite{sf2} was born in the same line of Kim and Markowitz model, the Black Monday and the consequent unexplainable worldwide crash in 1987. We can say that 1987 crash is a cornerstone which begun to erode the consensus on the standard efficient and rational financial theory in the economical community. \\\\
We briefly report the justification of the introduction of this model. The question of whether the market is efficient and rational or not is not only an interesting academic problem. In fact the conclusion of standard financial markets is that homogeneous rational agents endowed with the same information have no incentive to trade and the exchanged volumes should be nearly zero. In addition technical trading strategies cannot be profitable in an efficient world where prices always follow their fundamental value. It is curious that this vision of financial market was not seriously questioned until 1987 and it is also curious that the academic community of Economics does not recognize immediately that there was something wrong in standard financial theory. The perplexity of the authors of the present review is motivated by the fact that empirical evidences of large volumes, of persistent volatility, of speculative bubbles, of sudden crashes, etc.\\
On this account a very interesting discussion about the main differences between Economics and Physics with respect to their approach to empirical evidences can be found in \cite{libropatterns}. It is clearly reported that there is a systematic lack of empirical validation of economical models until very recent time. \\
In the last years changes have been observed and the attention to data is now becoming very important and we think that physicists can give a substantial contribution in turning financial economy in an observational science in which the criterion to refuse or accept a model can be only set through a validation with respect to empirical evidences. \\\\
\textbf{This model aims at investigating the regime in which a market tends when the agents' forecasting strategies are heterogeneous}. It is worth noticing that the authors maintain a neoclassical framework to model the market and the only new ingredient is indeed the heterogeneity in agents' forecasting action. Consequently the agents' homogeneous expectation does not hold and the price equilibrium problem cannot be anymore solved by deduction or by logical reasoning. In such a heterogeneous scenario this work wants to investigate whether or not the agents' strategies tend to a rational and homogeneous state or whether agents are able to select technical strategies and build a realistic dynamics of prices. 
\subsubsection{\textbf{Agents and Strategies}}
This artificial market is composed of two tradable financial assets, a risk-free bond with interest rate $r$ and a risky asset which pays a stochastic dividend $d_t$.\\\textbf{The set-up of the model is neoclassical}, i.e. the model is built following the framework of standard financial theory. Differently from other ABM this model has a micro-economical foundation in some extent. In fact agents aim at maximizing their utility function $U(c)$ where $c$ is their wealth. The authors choice a standard utility function $U(c)=-\exp(-\lambda c)$ with a constant absolute risk aversion whose is measured by $\lambda$ that is the same for all agents.\\
The number of shares of the risky asset is equal to the number of agents.\\
The maximization of the utility function corresponds to find the optimal composition of the portfolio between the bond and the risky stock. It is a well-known result of standard economical theory that the optimal allocation $x_{i,t}$ of the risky asset in such a scenario is
\begin{equation}\label{stratesfi}
x_{i,t}=\frac{E_{i,t}[p_{t+1}+d_{t+1}]-p_t(1+r)}{\lambda \sigma^2_{i,t,p+d}}
\end{equation}
where $E_{i,t}[\cdot]$ is the forecast function used by agent $i$ at time $t$ and $\sigma^2_{i,t,p+d}$ measures the confidence of the prediction $E_{i,t}[p_{t+1}+d_{t+1}]$. \\\\
\textbf{The deviation from the standard neoclassical scenario is introduced at this step}: the rationality and homogeneity hypothesis would be normally invoked, so that the agents' index can be neglected and strategies confidences are necessarily all equal. Consequently the agents can deductively find the solution of the clearing price equation and they would find that $p_t\sim\sum_{j=1}^{\infty}\beta^kE[d_{t+k}|I_t]$ where $\beta$ is a discount factor. In this way the process will reflect the price of the fundamental (i.e. the dividend) on average and the fluctuations of the price are only driven by the fluctuations of news $I_t$ (see \cite{rational}).\\
\textbf{The novelty of the model is the introduction of the heterogeneity in the agents' expectations} $E_{i,t}[\cdot]$. Now it is impossible to solve the clearing price equation by deduction since agents must guess the others' expectation rules and there are not any logical criterions to perform such a task. Then agents inductively learn, they have to compete in a heterogenous world of strategies and form their expectations by adaptation. 
\\\\In practice agents have $M$ expectational models called predictors. Each predictor is composed of the expectational strategy and of a part which determines in which state of the market the strategy is potentially available for the agents. The state of the market is described by a vector of $J$ bits at each time, if the corresponding part of the predictor matches the current indicator then the associated strategy can be used by the agent. If more than one strategy matches the current state of the market, the agent picks the one with the best score. In  fact when a strategy is used and when the clearing price is made public its score is updated and receives a higher score, the higher is its accuracy in the price forecast. \\In details the update of the accuracy of a strategy is the inverse of the moving average of the squared errors $e^2_{t}=(1-\alpha)e^2_{t-1}+   \alpha(p_{t+1}+d_{t+1}-E_{t}[p_{t+1}+d_{t+1}])^2$. The parameter $\alpha$ plays the role of the speed update of the quality of the strategy, the higher is $\alpha$, the higher is the weight of the last update. This is the first mechanism of selection of strategies: the most used and the most accurate strategies will be selected and used by the agents.
\\Before turning our attention to the second source of adaptation and selection we must describe how the expectational part of the predictor $E[\cdot]$ is made. The expectation function is linear and its parametrization is: $E[p_{t+1}+d_{t+1}]=a(p_t+d_t)+b $. Therefore in the end each predictor turn to be a strategy described by the couple of parameters $(a,b)$.\\
The second source is a mechanism which introduces new predictors discarding the worst performing strategies of agents. This happens through a genetic algorithm which asynchronously replaces the worst strategies with new ones obtaining them by picking and recombining two predictors of the agents.\\ \textbf{We are going to see that the frequence of activation of this algorithm is a crucial parameter for the model and will give rise to different regimes.}
\subsubsection{\textbf{Number of agents}} The number of active agents is constant over the time because all agents are endowed with the strategy which matches all the possible states of the market, so that there is always at least one predictor which matches the market state at each time step. The typical number of agents in \cite{sf2} is $25$. We are not aware of systematic study of the model properties with respect to the number of agents.
\subsubsection{\textbf{Price formation}} As in most ABM the feedback between agents' actions and prices is represented by the price clearing mechanism. The clearing price $p_t$ at time $t$ is the equilibrium price from the direct solution of the equation
\begin{equation}
\sum_{i=1}^Nx_{i,t} =N
\end{equation}
where $i$ is the agent's index, $x_{i,t}$ is the demand of shares of the agent $i$ at time $t$ ($x_{i,t}$ can be both positive and negative) and $N$ is the available number of shares .\\
At the beginning of the time step the value of the stochastic dividend $d_t$ is made public. The dividend follows an autoregressive process $d_t=\bar{d}+\rho(d_{t-1}-\bar{d})+\epsilon_t$, where $\bar{d}$ represents the average dividend paid by the risky asset, $\epsilon$ is a gaussian random variable and $\rho$ is the strength of the mean-reverting tendency of the autoregressive process. This parameter is chosen close to $1$ so that $d_t$ performs a quasi pure random walk. The explicit form of $x_{i,t}$ is given by eq. \ref{stratesfi}.
\subsubsection{\textbf{Origin of SF}} The model is able to reproduce both \textbf{the regimes} it was introduced for \textbf{ depending on the values of the parameter that rules the frequency of activation of the genetic algorithm}.\\
If this frequency is high enough the model reproduces a non rational regime - with respect to the concept of rationality in standard financial theories - otherwise the agents converge to a homogeneous strategy very close to the rational solution of this neoclassical artificial market and the price behavior follows the fundamental and small trade volumes are observed. The non rational regime is observed if the genetic algorithm is invoked each $250$ time steps, while the rational one if invoked each $1000$ time steps. The update of the strategies is asynchronously performed by the agents. \\\\
We focus our attention on the non rational regime which is characterized by \textbf{clustered volatility and fat tails}. Unfortunately the authors do not give a quantitative description of this SF. They generically say that fat tails are similar to the daily returns of IBM stock and a GARCH-like behavior of the volatility is observed. \textbf{The most interesting result is the emergence of technical strategies differently from the rational behavior}. In fact all predictors are divided into two groups, one part of bits measures fundamental properties of the price and the second part essentially are technical proxies for the price trends. Therefore it is possible to monitor whether fundamental or technical strategies are used by agents. Only in the non rational regime both strategies are selected and coexist. \\\\
We have said that the main parameter that determines which regime is observed is the frequency of activation of the genetic algorithm. If agents slowly explore the strategy phase space the rational homogeneous solution is an attractor. Instead if the rate of introduction of new predictors in the model is high enough the agents are not able to learn the rational solution and technical trading strategies self-reinforce their feedback so that they persist in this virtual market. However, the conclusion is in some extent trivial because \textbf{this frequency of activation plays the role of a source of disorder that can be seen as a sort of market temperature}. \\
For increasing values of this \textit{temperature} the strategy fluctuations introduced by the genetic algorithm do not give enough time to allow the emergence of the rational solution and instead these fluctuations permit the stable existence of technical strategies. In  this respect we argue that the real parameter that rules the regimes of the model is the ratio between the frequency of the genetic algorithm and the coefficient $\alpha$ which sets the speed of updates of the strategies. \\\\
In this interpretation the real stable solution is the rational one in agreement with the standard financial theory and the non rational regime is an artifact of the high rate of introduction of new strategies which prevents the learning and the adaptation of the agents. Obviously it can be argued that this could a possible explanation applicable to real markets too, where agents are not usually aware of other agents' strategies and of the changes of others' strategies. However we think that the explanation of the rich market ecology cannot be reduced to a system only driven by the uncertainty of the strategies operating in that system.
\subsubsection{\textbf{Discussion realism/tractability}} 
This is one of the few model with a \textbf{micro-economical foundation} and a set-up very close to the traditional approach to financial markets. \\
However, \textbf{by simply introducing the heterogeneity} in the future expectations of agents the model shows that \textbf{non-rational scenario dominated by technical traders can become the inductive} \textit{rational} \textbf{solution}, where by rational we mean that the technical strategies in such a  regime become competitive with respect to fundamental ones and are not eliminated from the market. \\The attempt to be a \textit{trait d'union} with classical economical theory is an element of realism. \\The model does not allow for any analytical approach and has a very limited tractability. In fact the model is very complicated nonetheless the two regimes appear to be ruled only by a single parameter. This suggests that this is not the minimal set up as for Kim and Markowitz model. \\In the end the main conclusion is that SF can be interpreted as a reinforced feedback effect mainly due to the strategy disorder. This conclusion would be the opposite opinion of many specialists according to which the coherence of strategies and their homogeneity can give rise to systemic instabilities and markets crashes.
\subsubsection{\textbf{Self-Organized Criticality}} The model shows that there exist two stable states which are mainly driven by the frequency of activation of the genetic algorithm. One of these two states produces a price dynamics which has statistical properties which are very similar to the real ones. Thus, as many other ABM, it shows that a stable and self-regulated state which exhibits SF can exist but the model does not address the problem of how this state can be reached and why markets should self-organize around this values of the frequency for the genetic algorithm. \\In brief the model does not cope with the dynamics of the market self-organization.  
\subsection{\textbf{Minority Game}}

\subsubsection{\textbf{Aim}}

Thanks to its mathematical structure, the Minority Game (MG) \cite{Challet:2005:MGI:1121663,Andrea2006:0305-4470:R465} is the most studied ABM. Consequently, many different versions of this model exist, with very different aims. The very first formulation, due to Arthur \cite{bar}, is known as ``The El Farol Bar Problem'' and \textbf{studies inductive reasoning in scenarios of bounded rationality}. \\Challet and Zhang \cite{mingame} gave a mathematical formulation of the problem. In this review we will analyze a model proposed by Challet et al. \cite{minstil} that specifically addresses the problem of the appearance of the SF, still conserving all the main ingredients of the classic MG. 

\subsubsection{\textbf{Agents and strategies}}

The MG is based on a simple concept. Let us suppose that \textbf{an agent can choose between two options} (these options are typically buy or sell a given amount of stocks). That agent will receive her payoff if, after all the agents choice, it will turn out that she took the minority side, that is she pick the less chosen option. This criterion may appear simplistic, but gets the point that one can exploit the price variations due to other agents' operations if one has previously taken the opposite side.\\
We now discuss the generic features of the MG.\\
There are $N$ agents in the market. At each time step $t$ each agent can choose to buy or sell stocks. Consequently, the function that describes the action of the i-th agent, $a_{i}(t)$, can be equal to either $+1$ or $-1$. We define the excess demand at time $t$ as
\begin{equation}
A(t)=\sum_{i=1}^{N} a_i(t). 
\label{hkhkgjgkg}
\end{equation}
It should be noticed that the agent can not choose the volume of her operation. \\\\
The fact that \textbf{the minority side wins} is considered introducing an utility function or a payoff given by
\[
U_i(t)=-a_i(t)\,F[A(t)]
\]
where $F$ is an odd function of $A(t)$. \\
Numerical simulations \cite{savit} have shown that the main features of the dynamics are left unchanged if one varies the functional form of $F$. In the following we will take $F[A(t)]=A(t)$.\\\\
At each time step, the public information is given by a random number $\mu(t)=1,...,P$ and the series of the previous $M$ results of the game: $+1$ if the majority has played $a(t)=1$ and $-1$ otherwise. So the parameter $M$ can be seen as the agents' memory, and the number of possible signals is $2^M$. \\\\
So far we have presented the generic rules of practically every version of the MG. Now we discuss the specific features of the model proposed by Challet et al \cite{minstil}.\\ The $N$ agents are divided in two groups with different strategies: the \textit{producers} and the \textit{speculators}. \\The former act in a deterministic way with respect to the state of the world $\mu$: the action $a_i(t)=\pm 1$  depends only on $\mu$ and it is extracted from a bimodal distribution with equal probability (any distribution with zero average and unit variance leads to the same results in the limit $N\rightarrow\infty$). So, \textbf{the producers inject information into the market}, representing the underlying economic activity and so  \textbf{ are similar to the fundamentalists in the Lux-Marchesi model} (see sec. \ref{lmsec}). \\ \textbf{ The speculators use adaptive strategies}, like in most versions of the MG, but with the possibility to not enter in the market. Each speculator has a set of strategies, that is maps from the available information (the state of the world and the previous outcomes) to the action, that can be buy, sell (that is, be active) or stay inactive.\\ In order to decide which strategy is adopt at time $t$, the agent estimates the performances of her strategies and chooses the best one. At each time step the performance of the strategy $s$ is given by
\[ 
U_{i,s}(t+1)=U_{i,s}(t)-a_{i,s}^{\mu(t)}A(t)+\epsilon\delta_{s_i(t),0}.
\]
where $\delta_{s_i(t),0}$ is Kronecker delta function.
Therefore, the active strategies are rewarded with the profit they would had brought if they have played, while the inactive strategy is rewarded with a constant premium $\epsilon$. Hence, an agent will enter in the market only if she had a strategy with an average gain larger than $\epsilon$. This can be easily interpreted as a risk-free asset such as a Treasury bill.
\subsubsection{\textbf{Number of agents}}
As we have seen in the previous section, the total number of agents is fixed, but at each time step they can choose to operate or not, so that  \textbf{ the effective number of active agents varies over time}. For this reason, using an analogy with Statistical Mechanics, one often refers to this model as the Grand Canonical Minority Game (GCMG).
\subsubsection{\textbf{Price formation}}
Being the excess demand $A(t)$ given by eq.(\ref{hkhkgjgkg}), the price is updated with the rule
\[
 \log p(t+1)=\log p(t)+\frac{A(t)}{\lambda}
\]
where $\lambda$ is a measure of the market depth. The return $r(t)$ is therefore basically given by the excess demand.  
\subsubsection{ \textbf{ SF and their origin}}
If only producers were active in the market the price would follow a random walk and the emergence of  \textbf{ the SF is due to the activity of the speculators}. Moreover, volatility clustering, fat tails and market crashes display a power law behavior and increase with the normalized number of speculators, $n_s=N_s/P$. The exponents depend on the specific parameters used in the simulation. The responsible for large price movements are the inactive agents that does not interact when out of the market but can enter at the same time if their strategies are similar.\\
Challet et al. \cite{PhysRevE.68.036132} studied a version of the GCMG that is very close to the one presented here. They found that in their model  \textbf{ the SF are finite size effects}, being present in a limited region of the parameter space that shrinks as $N^{-1/4}$ in the limit $N\rightarrow\infty$.\\
\subsubsection{\textbf{Realism and tractability}}
The level of realism and tractability is different for different versions of the MG. \\
Many realistic features have been added and studied taking the classic MG as a starting point: agents with finite or evolving wealth, agents using correlated strategies or acting at different frequencies, markets with more than one asset and so on (for these and other possible extensions, see \cite{Challet:2005:MGI:1121663,Andrea2006:0305-4470:R465} and references therein). If one considers all versions, many different aspects of real financial markets are taken in consideration.\\\\
The MG is, among all the ABM, the most studied from an analytical point of view. The first step is due to Challet et al. \cite{challet:marsili:zecchina}, that showed that the stationary state of the dynamics is related to the minima of the so called \textit{predictability}
\[
H=\overline{<A>^2}=\frac 1P \sum_{\mu=1}^{P}<A|\mu>^2.
\]
This observation opened a connection with Statistical Mechanics, as it turned out that the ground state was the same of a system of disordered spins. For example, using the replica method \cite{citeulike:1398137} they solved the basic model exactly and showed that the ground state is not a Nash equilibrium  \cite{kreps}, that is, it is not individually optimal. This kind of approach can give revealing insights into almost every version of the MG.
\subsubsection{\textbf{Self-Organized Criticality}}
The problem of Self-Organization is mentioned but not systematically addressed.

\subsection{\textbf{Caldarelli, Marsili and Zhang: a prototype of stock exchange}}
\subsubsection{\textbf{Aim}}
In this work \cite{caldase} the authors focus their attention on the endogenous mechanisms of financial markets and show that these ingredients are sufficient to obtain a stable and self-organized market in the sense that external (i.e. exogenous) rules are not required to stabilize the price behavior. 
\subsubsection{\textbf{Agents and Strategies}}
There are two tradable assets: a risk free asset with zero-interest rate (i.e. cash) and a risky stock. All agents are speculators so that they all aim at increasing their wealth and at each time time their action consists in updating their portfolio composition of stocks and cash according to their current strategy.
As in many models with two assets, the agent wealth is $w_t=M_t+p_tS_t$ where $M_t$ is the cash owned by the agent at time $t$, $S_t$ is the number of shares of the stock and $p_t$ is the price of the stock. \\
All the agents share the same information which consists in the price time series. \\The agents' strategies depend only on the price history and consists in updating her owned shares in the following way
\begin{equation}\label{strateMarsili}
\Delta S_{i,t} = X_{i,t} S_{i,t}+\frac{\gamma_iM_{i,t}-p_tS_{i,t}}{2\tau_i}
\end{equation}
where the index $i$ denotes the agents. The parameters $\gamma,\tau$ specify part of the agents' strategies. The speculative fraction $X_{i,t}$ of stocks $S_{i,t}$ which the agent would like to buy or sell (according to the sign) is
\begin{equation}
X_{i,t}=F_i[p_t,p_{t-1},\ldots]
\end{equation}
where $F_i[\cdot]$ is the strategy of agent $i$. In detail the function $F_i[\cdot]$ is a linear combination of moving time averages of derivatives of $\log p_t$
\begin{equation}
X_{i,t}=F_i\bigg[\sum_{k=1}^l \eta_{i,k}\langle \partial^k\log p_t\rangle\bigg].
\end{equation}
The choice of this kind of parametrization and the value of $l$ are somewhat arbitrary (on this account see sec. \ref{socmarsili}). On the other hand, a similar freedom is not allowed for the choice of $F_i[\cdot]$. In fact this function must satisfy the constraint $F_i[x]<1$ $\forall x$ and must be consistent with a realistic investment behavior. For instance for large crash agents could become very prudent, hence $F_i[x] \ll 1$ for  large values of $x$. This hypothesis may be questionable since large negative fluctuations can give rise to self-reinforcing waves of selling which cause the collapsing of financial bubbles. However, the final choice of the authors is $F[x]=x/[1+(x/2)^4]$. \\
 \\Let us now interpret the two terms eq. \ref{strateMarsili} is composed of. The first term on the right side of eq. \ref{strateMarsili} mimics a chartist strategy: in fact the speculative action of the agents is the result of the observation and the analysis of the past trends. Agents primarily act as technical traders. As all trend-based strategies, \textbf{their impact can be potentially destabilizing.} \\
The second contribution of the agent's action $\Delta S_{i,t}$ on the market plays a role similar to rebalancer agents in Kim and Markowitz model. In fact this contribution tends to rebalance the portfolio composition in order that $S_{i,t}/M_{i,t}=\gamma_i/p_t$. The parameter $\tau$ sets the characteristic time that is required to perform the portfolio rebalancing. \textbf{This term has a stabilizing }effect because the higher will be the price, the lower will be the relative exposure on the risky stock (relative with respect to the cash amount $M_{i,t}$). \\\\
To sum up, once fixed the parametrization of the strategy and the function $F_i[\cdot]$, the strategy of the agent $i$ is univocally determined by the set of parameters $\{\eta_1,\ldots,\eta_l,\tau,\gamma\}_i$. Thus, in this model agents are really heterogeneous with respect to the strategy, because the same price history leads to different actions differently from Kim and Markowitz model, where different decisions are the result of different signals and not of different strategies within the same class of agents. \\
\subsubsection{\textbf{Number of agents}}
While the total number is kept constant, the number of active agents are not a static quantity differently from Kim and Markowitz model. In fact at each time step the poorest agent is removed from the market and replaced by a new agent endowed with new random strategies. It is sense the population of traders is dynamic.
\\However, the question of the non-stationarity of the number of active traders is not addressed in the model.  
\subsubsection{\textbf{Price formation}} The feedback between prices and agents' actions is performed by a Walrasian mechanism of price update, that is the price response is proportional to the demand/supply imbalance. 
\\At each time step $t$ the previous clearing price is made public and each agent places an order according to eq. \ref{strateMarsili}. The excess demand $ED_t$ is the sum of the demand $D_t=\sum_{i,\Delta S_{i,t}>0}\Delta S_{i,t}$ and the supply $O_t=\sum_{i,\Delta S_{i,t}<0}\Delta S_{i,t}$. The lack of liquidity is introduced simply comparing supply and demand, for instance if $D_t > |O_t|$ then the fraction of stock that an agent is really able to buy is $\delta S_{i,t}=\Delta S_{i,t} O_t/D_t$ and viceversa if supply exceeds demand. The trading rules of the model are also provided with small buying transaction costs and small random fluctuations of the amount of agents' cash. The next price is fixed in order to be the price at which the excess demand is cleared and therefore
\begin{equation}
p_{t+1}=p_t\frac{\langle D_t\rangle}{\langle O_t\rangle}
\end{equation}
where $\langle \cdot \rangle$ is an exponential moving average. This equation describes a multiplicative evolution of prices and this is why authors choose the logarithmic of the price as the price indicator.
\subsubsection{\textbf{Origin of SF}} We can argue that \textbf{the competitive action} which is present at each time step in agent's operations $\Delta S_{i,t}$ \textbf{is the origin of the SF in this model}. In fact, if the model is suitably tuned, the two terms of eq. \ref{strateMarsili} counterbalance each other giving rise to a market which shows fluctuations at all scales without the need of exogenous regulatory rules to stabilize the market behavior. This is probably the origin of the \textbf{fat-tailed pdf of returns} that decays with an exponent close to $2$. This SF is not reproduced in a quantitative way since real markets show a larger exponent typically ranging from $3$ to $5$. \\
Moreover \textbf{the model does not have simple arbitrage opportunities} because the autocorrelation function of returns is substantially zero. In the end the return distribution shows a scaling exponent similar to the real one when the aggregation of returns is performed that is $F(r,\tau)=\tau^{0.62}g(r\tau^{-0.62})$ where $\tau$ is the time lag at which returns $r$ are calculated. The crossover towards gaussianity for large values of $\tau$ is not observed but this could be caused by the fact that all agents have a very similar time horizon of their investment strategies (in fact there is only a small degree of heterogeneity in the characteristic time $\tau$ of the rebalancing term of their strategies).\\
For the rank-size rule of agents' wealth the authors find a power law behavior with exponent very close to $1$, as in real cases.
\subsubsection{\textbf{Discussion realism/tractability}} Tractability is not a strong point of model. In fact the model requires a very fine tuning of strategies $F[\cdot]$ and of parameters in order to obtain a stable regime for prices, at least for the time scale investigated in the original paper ($10^3-10^4$ time steps. \\In \cite{calda_critica} it is argued that the model is asymptotically unstable that is the price converges to zero or diverges for $t\rightarrow\infty$). In this sense this model would not be really self-organized and would not be stable without external regulatory rules. \\However, the time unit of the model is not explicitly calibrated thus the model could be substantially stable for times long enough to conclude that endogenous ingredients are sufficient to reproduce a self-regulating market. \\Anyhow, this kind of instabilities is typical of multiplicative dynamics such as $p_{t+1}=\alpha_t p_t$, where even very small biases with respect to $\bar{\alpha}_t=1$ can lead to dramatic crashes or divergences at very long times. This can be explained by the fact that the functional form $F[\cdot]$ is fixed: then this could introduce a systematic bias.\\
Furthermore \textbf{multiplicative dynamics usually makes the tuning operation of the model very difficult} because multiplicative models does not even exists in a stable form in most of the phase space of parameters, and in this case in the phase space of strategies $F[\cdot]$ too. All these elements are conflicting with the aim of the authors which consider their prototype as a first step towards a systematic analysis of the role of each element of financial markets and, ultimately, with the tractability of this model.
\\\\With respect to realism, this model is one of the first ABM which is able to partially reproduce the main SF in a quantitative way and has the merit to reach this goal introducing only a \textbf{self-regulated market} through the feedback between price evolution and agents' actions. To our knowledge it is perhaps the first model which underlines the non-trivial problem of a self-organized market. The model shows that a quasi-stable self-organized state can exist but does not answer the question of how this state is reached and how it is restored.\\ It is worth noticing that the problem of a self-organizing market is neglected by economical literature. \\On the other hand complex systems usually exhibit self-organized or adaptive behaviors and therefore physicists are familiar with this kind of phenomena and the self-organization is consequently a problem which is typically handled in works by physicists (secs. \ref{giarbou} and \ref{nostro})
\subsubsection{\textbf{Self-Organized Criticality}}\label{socmarsili} This work is one of the first which deal with the self-organization of the market. The model produces a stable market given some particular classes of functions $F[p_t\,p_{t-1},\ldots]$ and some very restricted areas of the phase space of its parameters. This is not a surprising aspect since most of ABM require a fine tuning of parameters to obtain a realistic behavior and in some case even to obtain a stable behavior of prices. The authors argue that the simulated market is self-organized in the sense that external rules (i.e. exogenous forces) are not needed in order to obtain a realistic and stable market once the model is tuned. Therefore one of the most interesting conclusion of the model is that \textbf{a realistic price dynamics, that is a price dynamics which exhibits SF, can be explained by fully endogenous mechanism between agents' actions/strategies and prices}. \\
 But the key problem of market self-organization is not addressed. In fact why should markets be self-organized in this particular state deriving from the fine tuning of the model. Why only a particular class of strategies can give rise to a stable and realistic market and consequently why agents should select these strategies. They only show that a self-organized state can exist and can be generated by endogenous dynamics but they do not answer to the dynamics question of how the market reaches this self-organized state and how the market is able to restore this state if it is driven far from the state of SF. \\
 This point is usually neglected by most of ABM but some exceptions are the model reviewed in sec. \ref{nostro}. 
\subsection{\textbf{Lux and Marchesi model }}\label{lmsec}

\subsubsection{\textbf{Aim}}

The aim of the Lux-Marchesi model \cite{LMnature,LM} is to show that the scaling laws that are observed in financial markets can arise from agents' mutual interaction. In particular, Lux and Marchesi stress the clear difference between the statistical properties of the model input, i.e. the normal noise that makes the stock fundamental price evolve in time, and the output, that is the price dynamics produced by the agents' operations.

\subsubsection{\textbf{Agents and strategies}}

The agents are divided in two categories: the \textbf{ fundamentalists}, who believe in the existence of a fair price $p_f$ for the traded stocks (and so, they sell stocks if $p(t)>p_f$ and they buy if $p(t)<p_f$), and \textbf{the chartists}, that are noise traders whose behavior is dictated by herding and historical prices. While the total number of agents $N$ remains fixed throughout the simulation, the number of fundamentalists and chartists, $n_f$ and $n_c$, are allowed to vary.\\ The effect of the two classes of agents on the price is very different: while \textbf{fundamentalists have a stabilizing effect} on the market, as their operations drive the price towards the fundamental one, \textbf{chartists have a destabilizing effect and create bubbles and crashes}. Furthermore, chartists agents are divided in two subcategories, optimists, who believe that the price will rise and hence always buy stocks, and pessimists, who believe that the price will decrease and so, on the contrary, always sell stocks. Obviously, $n_+(t)+n_-(t)=n_c(t)$.\\
At each time step each agent can change her category with a given transition probability. In this way, an internal dynamics within agents' classes is established.\\ Using the same notation of \cite{LM}, the transition probabilities between fundamentalists and optimists and fundamentalists and pessimists are given respectively by
\begin{equation}
\pi_{+f}=\nu_2\bigg(\frac{n_+}{N}\exp(U_{21}) \bigg) \qquad \pi_{f+}=\nu_2\bigg(\frac{n_f}{N}\exp(-U_{21}) \bigg)
 \label{rateuno}
\end{equation}
and 
\begin{equation}
 \pi_{-f}=\nu_2\bigg(\frac{n_-}{N}\exp(U_{22}) \bigg) \qquad \pi_{f-}=\nu_2\bigg(\frac{n_f}{N}\exp(-U_{22}) \bigg).
\label{ratedue}
\end{equation}
The parameter $\nu_2$ is a constant and $U_{21}$ and $U_{22}$ are a sort of utility functions that quantify the difference between the instantaneous profits of the two strategies. The presence of the herding factor $\frac{n_x}{N}$ introduces an intermittent dynamics
between the populations and, as we will see in sec. \ref{duedopo}, is intimately connected to the emergence of the SF.\\
The transition probabilities between optimists and pessimists are
\begin{equation}
\pi_{+-}=\nu_1\bigg(\frac{n_c}{N}\exp(U_{1}) \bigg) \qquad \pi_{-+}=\nu_1\bigg(\frac{n_c}{N}\exp(-U_{1}) \bigg)
\label{ratetre}
\end{equation}
where again $\nu_1$ is a constant. Note that the herding term in eq. \ref{ratetre}) is different from the ones in eqs. (\ref{ratedue}) and (\ref{rateuno}): this difference will turn to be important to generate the SF.\\
The dynamics admits fixed points corresponding to zero agents in one of the categories. These cases are avoided by imposing suitable lower limits. 

\subsubsection{\textbf{Price formation}}
At each time step the price is increased or decreased of a tick (one cent) with a probability given by 
\[
 \pi_{up}=\textrm{max}[0,\beta(ED)+\mu] \qquad \pi_{down}=-\textrm{min}[0,\beta(ED)+\mu]
\]
respectively, where $\mu$ is a white noise that takes random traders into account and $ED$ is the excess demand, that is, the difference between the aggregated operations on the market.\\ In the case of fundamentalists, $ED_f=n_f\gamma(p-p_f)$: as we have seen, they sell if the stock is overvalued and buy if it is undervalued with respect to the perceived fair price. The chartists, instead, buy or sell a fixed amount $t_c$ of stocks according as they are optimists or pessimists, and so $ED_c=(n_+-n_-)t_c$. The parameters $\gamma$ and $t_c$ are constant. The total excess demand is $ED=ED_c+ED_f$.
\subsubsection{\textbf{SF and their origin}}\label{duedopo}
The herding terms in the transition probabilities between fundamentalists and chartists give rise to an alternation between the two metastable states in which one population is predominant. This process has been studied for the first time by Kirman \cite{kirman} and then, in two asymmetric versions that are closer to the Lux-Marchesi dynamics, by Alfarano et al. in \cite{alfasi} and Alfi et al. in \cite{paperoI}. These simplified versions permit to solve analytically the contagion models proposed. \\
In the Lux-Marchesi model the most probable state is the one in which the majority of agents is fundamentalist. However, a stochastic fluctuation can trigger a bubble of chartism that can persist for a long time. During this period one finds a strong alternation between optimists and pessimists, because their rates are both proportional to $\frac{n_c}{N}$. Consequently, many buy and sell orders arrive in the market and the price goes quickly up and down accordingly, causing an increment of volatility. \textbf{This causes an intermittency between regimes which have fluctuations with different amplitudes and, consequently, the emergence of fat tails and volatility clustering}. \\
Zawadowski et al. in \cite{Zawadowski2002403} have studied the response to fundamental price drops within this model, finding a realistic overshoot.
\subsubsection{\textbf{Realism and tractability}}
Lux and Marchesi have included various reasonable aspects in their model: the importance of herding, the price evolution as a consequence of excess demand, the alternation of stabilizing and destabilizing factors in the market. In order to do this, they have built a complicated dynamics in which the roles played by the 13 parameters involved is not clear. This leads to a realistic dynamics, but reduces the tractability of the model. However, in \cite{LM} the properties of stationary states are analyzed. In these states the price is equal to the fundamental one and there are no net flows among the populations. The authors have studied the stability of these states and their relationships with the emergence of the SF.
\subsubsection{\textbf{Number of agents}}
In the Lux-Marchesi model the total number of agents $N$ is fixed. On the contrary, the number of agents in each class can vary. Egenter et al. in \cite{LMfinite} have shown that in the limit of large $N$ the Lux-Marchesi model is no more able to reproduce the SF. In this case all the agents become fundamentalists and so the price strictly follow the fundamental price. This mechanism can be easily understood if one considers a simplified asymmetric model for the population dynamics \cite{paperoI,paperoEPL,paperoII}.
\subsubsection{\textbf{Self-Organized Criticality}}
The problem of self-organization is not addressed in this paper.
\subsection{\textbf{Giardina and Bouchaud model}}\label{giarbou}
\subsubsection{\textbf{Aim}} The model introduced in \cite{giard_bou} is built to be, as explicitly declared by the authors, a \textit{trait d'union} between the tractability of Minority Games and the complicated Sante Fe virtual stock market. However, the resulting model is still rather complicated and a fine tuning of the model parameters is required. In fact this model is characterized by a large amount of parameters but the SF are only reproduced in a very small region of the parameter space. Then \textbf{the natural question is why markets should self-organize in this particular set of parameters which are able to reproduce SF} and how this state is restored by market dynamics if some fluctuations drive the system out of this state. The novelty of this model is that for the first time the problem of the self-organization is, at least qualitatively, discussed and addressed.  
\subsubsection{\textbf{Agents and Strategies}}
There are two tradable assets: a risky stock and a risk free bond with interest rate $r$. \\
As in Santa Fe artificial stock market agents inductively adapt their strategies and learn from their past performances. All agents share the same information, that is the price times series up to the present, and agents at each time step can revise their portfolio composition by selling, buying or not varying the number of owned shares. 
\\The update of the portfolio is determined by the strategy with the best score at that time. Let us describe how this general set up is practically realized in \cite{giard_bou}.\\
There are $N$ agents which have $S$ strategies of which $S-1$ produce an action on the market and one instead makes the agent inactive. Each agent has a portfolio described by the couple $(\phi(t),B(t))$, where $\phi(t)$ and $B(t)$ are the number of stocks and bonds respectively.\\ The wealth of an agent at time $t$ is $w(t)=\phi(t) p(t)+B(t)$ where $p(t)$ is the stock price.\\
The information shared by the agents are the last $m$ price values and in details the agents only \textit{remember} the signs of the returns. \\
The strategies can be \textit{polarized} at the beginning of the simulation, where for polarization we mean that the parameter $P\in[-1,1]$ takes values $P<0$ if contrarian are favored on average and $P>0$ if trend following strategies are favored. If $P=0$ there is not any bias in the strategy space. \\
\textbf{The agents are substantially chartists on a time scale $m$} since they look at the price behavior to determine their action. However, if they estimate that the price is very far from its fundamental value, the probability of activation of a fundamentalist strategy grows regardless of the score of the chartist strategies. In such a way a stabilizing tendency is introduced when prices are too high or too low with respect to the fundamental value. \\
On the other hand the effect of the chartists can either stabilizing or destabilizing depending on the nature of their strategies that can be, on average, trend-contrarian or trend-follower. \\
Nevertheless the nature of the strategy originating the decision, the agents can buy or sell stocks (or be inactive) converting a fixed fraction of their bonds in stocks or viceversa. This fraction $g$ is independent on the agent and is assumed equal to $\sim 1\%$. In formulas if an agent decides to buy, the order volume $q(t)$ is
\begin{equation}
q(t)=g\frac{B(t)}{p(t)}
\end{equation}
 otherwise 
 \begin{equation}
q(t) = -g\phi(t).
\end{equation}
If the agent stays inactive then $q(t)=0$. \\
In the end, agents can randomly decide if buying or selling $q(t)$ with a small probability independently on the price value or the fundamental value.  This strategy can be interpreted in terms of irrational agents or in terms of occurrence of errors in trading activities.
\subsubsection{\textbf{Number of agents}} 
The total number of agents is constant but \textbf{agents can decide to be active or inactive on the market}. Therefore the number of active agents changes along time but this number is bounded from above. In addition, in this model the number of active agents is substantially proportional to the trading volume. In section \ref{sf_bou} we analyze the role of the variability of the number of agents with respect to volatility clustering and intermittency. On this account the presence of the inactive strategy is crucial in order to reproduce the intermittent regime in which SF are observed. \\
The authors of this model investigated a wide range of $N$ up to $10000$ but they do not observe a significant dependence of the model features on $N$, differently from Lux and Marchesi model.
\subsubsection{\textbf{Price formation}}
At each time step the relative order imbalance $Q(t)$ is 
\begin{equation}
Q(t)=\frac{1}{\Phi}(q^+(t)-q^-(t))
\end{equation}
where $\Phi$ is the total number of shares of the risky stock which is assumed to be constant on time, so that $Q^+(t)=q^+(t)/\Phi$ and $Q^-(t)=q^-(t)/\Phi$ are  the fractional volume of buy and sell orders respectively. \\\\
The clearing mechanism simply assumes that the log returns of prices are proportional to the demand imbalance such that 
\begin{equation}
\log(p(t+1)/p(t))=\lambda^{-1}Q(t).
\end{equation}
The parameter $\lambda$ can be interpreted in terms of market stiffness: the larger is $\lambda$, the smaller is the market response given the same imbalance $Q(t)$. \\The authors also analyzed the effect of other choices for the clearing rule assuming $\log(p(t+1)/p(t))=\lambda^{-1}Q(t)^\alpha$ with $\alpha<1$ as in real market \cite{lillocoll,revBOU} but there is not a crucial dependence on the price update rule in this model. \\\\
In general the number of sell orders is not perfectly matched by buy orders and then the model introduces a phenomenological mechanism to ensure that the number of sold shares is equal to the number of bought ones. The number of shares that can be effectively bought at the new price is 
\begin{equation}
\bar{Q}^+(t)=Q^+(t)\frac{p(t)}{p(t+1)}
\end{equation}
and consequently the fraction of filled buy and sell orders are respectively $\phi^+=\min(1,Q^-/\bar{Q}^+)$ and $\phi^-=\min(1,\bar{Q}^+/Q^-)$.
The amount of riskless bond and risky stocks are updated according to $\phi^+$ and $\phi^-$. \\
The microscopic interpretation of such a clearing mechanism is that there is a fraction of orders that are usually unexecuted as it happens in a real order book with a double auction mechanism. In some extent this choice for the clearing mechanism corresponds to the existence of market orders, i.e orders immediately executed and limit orders, i.e. orders stored in the order waiting for a counter order able to match it. \\\\
The strategies and the agents' actions on the market set the price update. The price variation determines the effective portfolio evolution ($\phi^+,\phi^-$) and in the end the update of the scores of the strategies an agent is endowed with. All the strategies are updated, both if they have been used and if they are inactive at that time.  At each time the new score of a strategy is the weighted average of the previous values of the score and of the virtual profit of that strategy. The parameter that defines the weight of this average sets the memory on which the strategy performances are remembered by agents.   
\subsubsection{\textbf{Origin of SF}}\label{sf_bou}
The model is characterized by three regimes which depend mainly on a couple of parameters, the ratio $g/\lambda$ which fixes the market response to an imbalance and the polarization $P$ of the chartist strategies. \\
A \textit{rational} regime is observed for large negative polarization ($P$ close to $-1$) because the chartist strategies are almost all mean reverting and the stock price follows the fundamental one. \\
An oscillating regime is found for $g/\lambda <0.4$ and positive polarization, where periodic bubbles and successive crashes take place. This regime is characterized by large arbitrage opportunities and inefficiency.  \\
The most interesting regime which is obtained for $g/\lambda \geq0.4$ and $P\geq-|P_0|$ ($|P_0|\sim0.05$) reproduces most of the market SF and is almost efficient. \\\\We now focus our attention on this intermittent regime which exhibits realistic features.
\textbf{The price time series shows volatility clustering} (only for large values of $g/\lambda$) \textbf{and fat tails} (the exponent of the tail distribution of returns is around $3.5$).\\The investigation of the scaling properties of returns with respect to the time lag $\Delta t$ on which the return is performed reveal a Ornstein-Uhlenbeck-like behavior, i.e. the price is mean-reverting on long time horizon and consequently the price fluctuations become constant and independent of $\Delta t$ for large values of the time lag. This mean-reverting tendency is due to the finite wealth that cannot indefinitely fuel a divergence from the fundamental price. In fact the same behavior is observed when all the agents' strategies are random and  therefore this effect is only due to the constraint represented by the finite capital. Such a convergence to a constant value is not observed in real price time series.\\ The origin of the volatility clustering can be traced back to the fact that volume and volatility are correlated only in the high activity region while they result uncorrelated in low activity time. This non linear relationship produces a non trivial behavior for the volatility (see \cite{giard2} for further details). 
\subsubsection{\textbf{Discussion realism/tractability}}
The model matches most of the SF but, as Lux and Marchesi model, its building blocks are far from being the very minimal ones. Despite it can be approximately described by only two quantities $P$ and $g/\lambda$ and allows for some analytic approaches in some regimes, the model has never been used for a systematic investigation of the market properties. In addition the model does not integrate phenomenon such as herding or contagion which may be one of the principal mechanism which can fuel the growth of speculative bubbles. 
\subsubsection{\textbf{Self-Organized Criticality}}
To our knowledge this is the first work that really addresses the question of\textbf{ the dynamics of the self-organization of the markets towards the intermittent state of the SF}. Unfortunately the authors does not give an answer to this problem because they do not investigate the interesting argument proposed in \cite{giard_bou} in the framework of their model. In fact they observe that in principle for small values of the ratio $g/\lambda$ the model produces very long bubbles and a reasonable behavior for investors in such a scenario would consist in increasing the invested fraction $g$ (the model rules do not allow for such a dependence, $g$ is fixed along the simulation) and therefore $g/\lambda$ will spontaneously increase. We recall that the intermittent regime is observed for  $g/\lambda \geq0.4$. Meanwhile small values of $g/\lambda$ also produces small rate of executions and therefore the market will tend to be more liquid that is $\lambda$ will decrease. Both the effects contribute to enhance the ratio $g/\lambda$ and the market enter into the intermittent regime. Once this quasi efficient regime with burst of activity is reached the ratio should become almost constant because the quasi-efficiency and the burst of volatility does not fuel anymore the growth of $g/\lambda$. \\
In the end it is interesting that the intermittent regime is obtained when the chartists are not mean reverting on average. \textbf{This suggests that the destabilizing effect of trend following strategies is crucial to understand and reproduce the intermittency of real markets}. A similar conclusion is drawn in the model proposed in \cite{paperoI,paperoEPL,paperoII,paperoSOI}. 

\subsection{\textbf{The destabilizing effect of leverage}}

\subsubsection{\textbf{Aim}}

The aim of the model proposed by Thurner et al. in \cite{thurner} is to show how fat tails and volatility clustering can arise from the use of leverage and margin calls. This is done by running the simulations for different values of leverage.

\subsubsection{\textbf{Agents and strategies}}

Thurner et al. consider two types of traders: noise traders (nt) and hedge funds (h). Both can hold assets (for example, stocks) or cash, or a combination of the two. In addition, there is a representative investor who invests (withdraws) her cash in the funds if their performance exceeds (falls behind) a given benchmark return $r^{bm}$. We now describe the agents in detail.\\ Noise traders' operations are random but still with a bias in order to make the price dynamics a mean reverting random walk with expected value equal to the fundamental price $p_f$. The fundamental price is common for all the agents and held constant throughout the simulation. Their aggregate  demand is
\[
D_{nt}=\frac{\xi_{nt}(t)}{p(t)}
\]
and the time evolution of $\xi_{nt}(t)$ is 
\[
\log \xi_{nt}(t+1)=\rho \log \xi_{nt}(t)+\sigma \eta(t)+(1-\rho)\log (N_s p_f),
\]
where $\rho$ and $\sigma$ are fixed parameters, $\eta$ is extracted from a normal distribution and $N_s$ is the number of shares of the asset.\\ 
The $h$-th hedge funds' wealth is the value of the asset plus its available cash
\[
W_h(t)=D_h(t)p(t)+C(t)
\]
where $D_h(t)$ is the demand (in this model the agents' demand is always fulfilled). If the fund perceives a signal to buy, the cash is decreased accordingly. If $C(t)$ becomes negative, the fund takes out a loan from its bank with a certain leverage $\lambda_h(t)$ given by
\[
\lambda_h(t)=\frac{D_h(t)p(t)}{W_h(t)}.
\]
In any case, this leverage is upper bounded by $\lambda_h^{M}$, that, in general, can be different for different funds. However, the main features of the model are already present if $\lambda_h^{M}=\lambda^{M} \,\, \forall h$, so that we will restrict our analysis to this case.We can anticipate that $\lambda^{M}$ is the key parameter to reproduce the SF: in fact for $\lambda^{M}=1$ (that is, no leverage) the fat tails are no longer produced.\\\\
Now we turn on the traders' operations on the market. In order to decide to enter in the market, the funds evaluate the asset mispricing $m(t)=p_f-p(t)$ and compute their demand accordingly. If $m<0$, the asset is perceived as overpriced and so the demand is zero, that is the fund does not enter in the market. Otherwise, the funds want to buy and the demand is proportional to both the mispricing and the fund's wealth:
\[
D_h=\beta_h\frac{mW_h}{p}
\]  
where the parameter $\beta_h$ is a measure of the fund's aggressivity. For large mispricings the fund may need an amount of cash which exceeds the maximum leverage it can ask to the bank: in this case its demand will be set to the maximum value
\[
D_h^M=\lambda^M\frac{W_h}{p}.
\]
If the fund stays under the maximum leverage and the price of the asset decreases, the fund must sell some of its assets in order to pay off its debt. This is called a \textit{margin call}. \\
A fund can go bankrupt if its wealth falls down a fixed threshold ($10\%$ of its initial wealth $W_0$ in the simulations). In this case it is removed and reintroduced after $T_{wait}$ time steps. 

\subsubsection{\textbf{Price formation}}

Given the traders' demands, the asset price is obtained imposing that at each time step the quantity supplied equals the quantity demanded, that is
\[
D_{nt}+\sum_h D_{h}=N_s
\]
where the demands are explicit functions of the price $p(t)$, the sum is extended over all the hedge funds present in the market and $N_s$ is the number of shares of the asset.

\subsubsection{\textbf{SF and their origin}}

The authors have studied the statistical properties of price returns in two cases: the one in which only noise traders are present and the one in which funds are active too with different values of the maximum leverage $\lambda^{M}$.\\ With only noise traders the returns are (by construction) approximately distributed according to a gaussian. For $\lambda^{M}=1$, that is funds are present but cannot use leverage, again the distribution does not have fat tails. \\\\\textbf{Letting funds use leverage creates fat tails}. \textbf{The tails are very asymmetric} due to the fact that funds can only have long positions (because they act only when the mispricing is positive). \textbf{The tail of negative returns can be fitted with a power law} and the exponent decreases monotonically increasing $\lambda^{M}$. \\ 
\textbf{The autocorrelation function} of absolute returns for different values of $\lambda^{M}$ is also studied. It \textbf{decays with a power law} with an exponent between 0 and 1, as in real financial markets, even if no leverage is possible. It should be stressed that in this last case the correlation is reduced but this values are closer to the ones found empirically.\\
All these properties are computed only in those time steps in which funds are active (that is, when $m>0$).\\\\
The origin of the SF can be explained in the following way. When only noise traders are present and when the funds cannot use leverage the price returns are normally distributed. If the use of leverage is allowed and funds have not reached $\lambda^{M}$, volatility is reduced, because funds tend to balance the mispricing, buying when the price is falling and therefore damping price movements. Instead, funds amplify volatility as soon as the leverage reaches $\lambda^{M}$, because in this case the fund must sell when the price falls down because of the margin calls. Thus, a positive feedback is established, since the sales cause a further price drop, that in turn causes more sales and so on. These arguments can be made more quantitative under reasonable assumptions \cite{thurner}.
\subsubsection{\textbf{Realism and tractability}}
The model is focused on the investment dynamics of and on hedge funds and it is realistic and accurate in this aspect. Concepts like irrational behavior, herding and chartism are voluntarily not addressed, even if many authors suggest them as the principal causes of SF. The dynamics of the margin call is realistic. The authors, perhaps by chance as themselves admit, find a reasonable exponent for the power law fit of the return distribution for a value of the maximum leverage that is equal to the one allowed in the United States.\\
The high number of free parameters (about 15) prevents an analytical tractability of the model. Nevertheless, in the Supplementary Information the authors, using some reasonable approximations, derive an estimation of the volatility for both the regimes of low and high leverage.
\subsubsection{\textbf{Number of agents}}
Both the noise traders and the hedge fund investors are treated at an aggregated level and can be considered as representative agents. For example, the cash invested or withdrawn from a fund is the sum of all the cash flows from the traders. On the contrary, the number of funds ranges from 1 to 10. The number of active funds can vary according to the rules described in the previous sections.
\subsubsection{\textbf{Self-Organized Criticality}}
The problem of self-organization is not addressed in this paper. Most of the parameters are fixed in all the simulations. The key parameter for the emergence of the SF is $\lambda^{M}$, that is not varied during a single simulation.
\subsection{\textbf{Credit network and bankrupt avalanches}}

\subsubsection{\textbf{Aim}}

The aim of the model introduced by Delli Gatti et al. in \cite{RePEc:spr:jeicoo:v:4:y:2009:i:2:p:195-212} is to study the properties of a credit network and the causes of the emergence of bankruptcy avalanches.

\subsubsection{\textbf{Agents and strategies}}

In this model there are three populations of agents: $I$ downstream (D) firms, that produce consumption goods using labor and intermediate goods; $J$ upstream (U) firms, that produce intermediate goods on demand from D firms and $Z$ banks to whom firms may demand credit if their net worth is not enough to finance the wage bill.\\\\
The \textit{D firms} demand both for labor $N_i(t)$ and for intermediate goods $Q_i(t)$ depending on their financial conditions, that is captured by net worth $A_i(t)$. Respectively, the demands of the \textit{i}th firm are given by
\[
N_i(t)=c_1 A_i(t)^{\beta} \quad \textrm{and} \quad Q_i(t)=c_2 A_i(t)^{\beta}.  
\]
The consumption goods are sold at a stochastic price $u_i(t)$ that is a random variable extracted from an uniform distribution in the interval $(0,2)$. The justification of these choices can be found in \cite{RePEc:spr:jeicoo:v:4:y:2009:i:2:p:195-212}.\\\\
The \textit{U firms} produce intermediate goods employing only labor, so that for the \textit{j}th the demand is
\[
 Q_j(t)=c_3 N_j(t).
\]
The authors assume that many D firms can be linked to a single U firm but each D firm has only one supplier for intermediate goods among the U firms. The price of intermediate goods is
\[
 p_j(t)=1+r_j(t)=\alpha A_j(t)^{-\alpha}
\]
where $r_j(t)$ is the interest on trade credit, which is assumed to be dependent only on the financial condition of the U firm. In particular, if the \textit{j}th firm is not performing well, it will give credit with a less favorable interest rate.\\
While the production of D firms is determined by their worth $A_i(t)$, the production of U firms is determined by the demand on the part of D firms
\[
 Q_j(t)=c_2 \sum_{i \in \Phi_j} A_i(t)^{\beta}
\]
where $\Phi_j$ is the set of D firms that are customers of the \textit{j}th U firm. Analogously, the demand for labor will be
\[
 N_j(t)=c_1 \sum_{i \in \Phi_j} A_i(t)^{\beta}.
\]
\\
If the net worth of the firms is not sufficient to pay the wage bill $W=c_4 N$ they will demand credit to a \textit{bank} (we have not rewritten the $i$ and $j$ subscripts, as the previous formula may refer to a D or an U firm). For each firm the credit demand is
\[
 B(t)=c_4 N(t) - A(t)
\]
where the functional form of $N(t)$ changes if we are considering an U or a D firm. Again, we assume that many firms can be linked to a single bank but each firm has only one supplier of loans. Without entering in the details, we point out that the interest rate on loans
is a decreasing function of the bank's net worth and penalizes financially fragile firms.
  
\subsubsection{\textbf{Number of agents}}

The total number of agents and the number of banks, U and D firms are kept \textbf{constant} over time: bankrupt agents are replaced one to one. However, in order to avoid cycles, the initial capital of the new entrants are small with respect to the average net worth.

\subsubsection{\textbf{Description of the dynamics}}
Firstly we define the network dynamics, that is, how D firms look for its U firm and how D and U firms look for a bank to take their loans. In order to establish the credit links the agents take a \textit{preferred partner choice rule}: at each time step each customer searches for the minimum of the prices charged by a randomly selected set of possible suppliers. It can change supplier only if a better partner has been found.\\
The agents' profits $\pi$ are evaluated from the difference between their gains and the costs they have to face; for example, the \textit{j}th U firm at time $t$ has a profit given by
\[
 \pi_j(t)=(1+r_j(t))Q_j(t)-(1+r_j(t))B_j(t).
\]
At each time step the net worth of the generic agent is updated according to the rule
\[
 A(t+1)=A(t)+\pi(t)-D(t)
\]
where $D(t)$ is the so called ``bad debt'', that takes into account the possibility that a borrower cannot pay back the loan because it went bankrupt (that is, $A(t)\leq0$). In this framework the lenders are the U firms and the banks and both U and D firms can be borrowers.

\subsubsection{\textbf{SF and their origin}}

The main goal of the paper is the study of \textbf{bankruptcy avalanches and their connection with the dynamics of the credit network}. From this point of view, the comparison with data is difficult, so we limit ourselves to the description of the mechanism that leads to the emergence of such avalanches. \\Suppose that a random price fluctuation causes the bankruptcy of some U firms. Consequently, the loans they took will not be fulfilled and the worth of the lenders (banks and D firms) will decrease. Eventually, this will result in a bankruptcy of some of them and, more importantly, in an increase of the interest rates charged on their old and new borrowers. This, in turn, will increase the probability of a bankruptcy of a D firm, and so on. \textbf{The credit network has a scale free structure and then the default of a highly connected agent may provoke an avalanche of bankrupts}.\\
Regarding other empirical evidences, this model shows reasonable agreement with the known data: the distribution of firm size is a power law and the one of aggregate growth rates is a double exponential (see \cite{RePEc:spr:jeicoo:v:4:y:2009:i:2:p:195-212} and references therein).

\subsubsection{\textbf{Realism and tractability}}

This model displays quite realistic features, starting from the three-sector structure of the credit network. Also the bankruptcy avalanches mechanism is realistic. Moreover, the model is able to reproduce the main empirical facts about the statistical properties of firms. \\

\subsubsection{\textbf{Self-Organized Criticality}}

The problem of self-organization is not addressed in this work.
\section{\textbf{Stylized Facts as finite size effects and self-organization of markets}}\label{nostro}

\subsection{\textbf{Aim}}

The aim of the model introduced in \cite{paperoI,paperoEPL,paperoII,paperoSOI} is to look for the \textbf{essential requirements} that an ABM must satisfy in order to reproduce the SF. In this sense, this model is \textit{minimal}. Moreover, this approach permits to investigate the problem of the \textbf{self-organization of financial markets}, that is, to find out why the empirical time series display a power law behavior typical of a (quasi) critical state without having any exogenous fine tuning.

\subsection{\textbf{Agents and strategies}}

The model takes and revises four core assumptions from the Lux-Marchesi (LM) model: the idea of \textbf{fundamentalists and chartists}, the central role played \textbf{herding} and \textbf{the mechanism of price formation}. Here we discuss the first three points; the discussion about price formation can be found in sec.\ref{gfsgsg}.\\\\
There are two classes of agents: $N_f$ \textbf{fundamentalists}, whose action has a \textbf{stabilizing effect} on price dynamics, and $N_c=N-N_f$ \textbf{chartists}, who act as a \textbf{destabilizing component}.\\
As in the LM model, the fundamentalists share a common opinion about the fair value of the traded stock, that we call the fundamental price $p_f$ which is estimated independently of price behavior, for instance evaluating the health of a firm, its management, its competitors and so on. This kind of approach is known as fundamental analysis from which the name of the class of agents comes out.\\
The chartists, instead, use historical data to take decisions. More specifically, they use the moving average $p_M(t) \equiv \frac{1}{M} \sum_{i=0}^{M-1} p(t-i)$ as a reference. Unlike fundamentalists, however, chartists believe that future prices will move away from $p_M(t)$ and act consequently. As a result, they tend to follow trends, both up and downward.\\\\
At each time step \textbf{agents can change their mind} according to a probabilistic rule that takes into account two factors: a price signal, that is the distance between the actual price and the reference value, and a herding term that depends on the fraction of agents who belong to the other class in order to consider the agents'  tendency to imitation. In formulas, the rate equation for the transition from chartists to fundamentalists is  given by
\begin{eqnarray}
P_{cf}\propto(K+\frac{N_f}{N})\exp(\gamma |p_f-p|)
\label{eq:5}
\end{eqnarray}
and the rate from fundamentalists to chartists is
\begin{eqnarray}
P_{fc}\propto (K+\frac{N_c}{N})\exp(b \frac{ |p_M-p|}{M-1}).
\label{eq:6}
\end{eqnarray}
where $\gamma$ and $b$ measure the strength of the price signal and $K$ corresponds to the possibility of neglecting the opinion of the other agents if the signal is strong enough. Without such a term the dynamics would be trivially convergent towards the two absorbing states corresponding to all agents being fundamentalists or chartists. 
\subsection{\textbf{Population dynamics}}
\begin{center}
 \begin{figure}[ht]
\includegraphics[width=11cm]{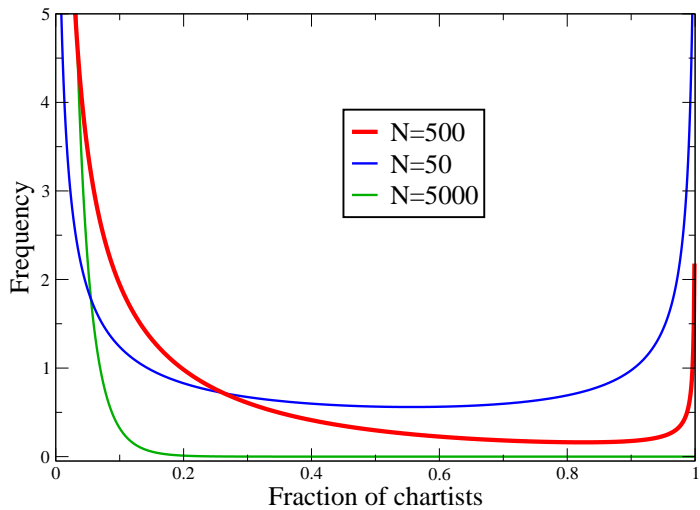}
\caption{\label{fig:1}
Stable asymptotic distributions for agents' strategies. In the left side fundamentalists dominate while chartists dominate on the right side. The model is slightly asymmetric in favor of 
fundamentalists, but this asymmetry depends on $N$. For a small number of agents the blue line
shows an almost symmetric distribution but
for larger values of $N$ fundamentalists become dominant. The specific values of $N$ depend on the other parameters of the model.}
\end{figure}
\end{center}

In order to derive an analytical expression for the stable distribution of $x=N_c/N$ the authors consider the following transition rates
\begin{eqnarray}
P_{cf}=B(1+\delta)(K+\frac{N_f}{N})\\
P_{fc}=B(1-\delta)(K+\frac{N_c}N)
\label{eq:8}
\nonumber
\end{eqnarray}
where $\delta$ is a parameter of asymmetry. This may appear a rough approximation but captures the essential elements of the dynamics and permits an analytical derivation of the asymptotic distribution of the fraction of chartists $x$, which turns out to be
\begin{eqnarray}
P_{eq}(x)\sim x^{r(1-\delta)-1}(1-x)^{r(1+\delta)-1}\exp{(-2\delta N x)}
\label{eq:9}
\end{eqnarray}
(see \cite{paperoI} for a detailed derivation). \\In fig.\ref{fig:1} we plot $P_{eq}(x)$ for different values of the total number of agents $N$ keeping all other parameters fixed. One can easily see that the asymmetry towards fundamentalists is greatly enhanced for large $N$: this is the very same phenomenon empirically shown by Egenter et al. for the LM model \cite{LMfinite}. 
\subsection{\textbf{Price formation}}
\label{gfsgsg}
The two classes of agents operate in the market (and thus affect the price dynamics) in a different way. As in the LM model, fundamentalists sell if they believe that the asset is overvalued, that is, if the stock price is above its fundamental value $p_f$; otherwise they buy.\\
Besides that, we want to make their contribution independent on the total number of agents. As a consequence, their demand is given by $\frac{N_f}{N}\gamma(p_f-p(t))$. Since chartists tend to move away the price from the moving average $p_M$, their contribution to the price dynamics is analogous and the complete price evolution equation is given by 
\begin{equation}
p(t+1)=p(t)+ \frac{N_c}{N}  \frac{b}{M-1}(p(t)-p_M(t))+\frac{N_f}{N}\gamma(p_f-p(t))+\sigma\xi
\label{eq:3}
\end{equation}
where $\sigma\xi$ is a white noise and the term $M-1$ makes the chartists' signal independent on the time window $M$ on which they evaluate their moving average \cite{taka1,alfi2,alfi3}.
\subsection{\textbf{SF and their origin}}
As we will see in the next sections, \textbf{the total number of agents $N$ is a crucial parameter in order to reproduce SF}. In this section we will focus on the origin of SF assuming that we already have the ``right'' number of agents.\\
As one can easily see from eq.(\ref{eq:8}), if $\delta > 0$ the most probable state is the one in which all agents are fundamentalists. This fact can be easily argued also from fig.\ref{fig:1}: only when $N$ becomes small the symmetry is reestablished. However, as in the LM model, \textbf{for finite values of $N$ bursts of chartism can sometimes occur}. In these periods the volatility increases and bubbles and crashes occur, until the fundamentalists become dominant again. \textbf{This mechanism causes an alternation of phases of low and high fluctuations} and, ultimately, \textbf{the emergence of fat tails and volatility clustering}. In order to obtain a more realistic relaxation for the autocorrelation function of absolute returns it turns out to be enough to introduce a realistic heterogeneity in the chartists' time scales $M$. 
\subsection{\textbf{Number of agents}}
\begin{figure}[ht]
\includegraphics[clip=true,keepaspectratio,width=11cm]{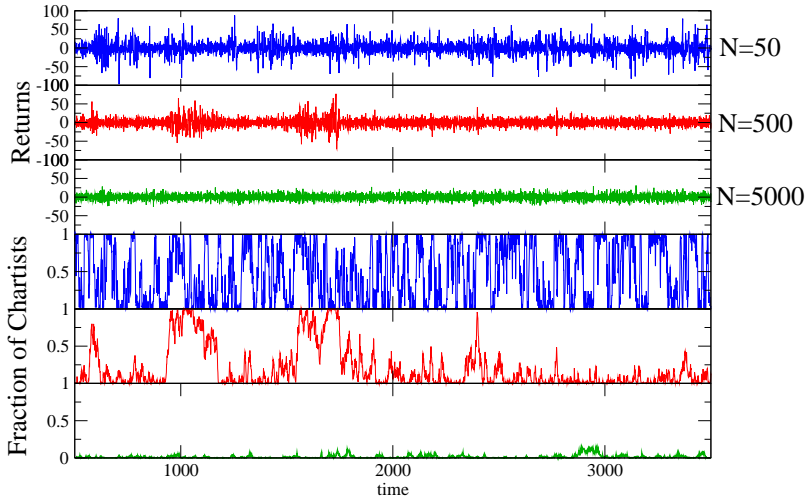}
\caption{\label{fig:2}
Dynamical evolution of the agents' strategies (bottom) for the three cases discussed in
fig.~\ref{fig:1} and the corresponding fluctuations 
of returns (top). One can see that a realistic
intermittent behavior of the returns 
is obtained only for $N=500$. This is due to the fact that for large $N$
$(5000)$ the system is stuck in
the fundamentalist state, while for smaller values of $N$ $(50)$ it fluctuates
too rapidly.
It is clear therefore that the
intermittent region corresponds to finite size effects which disappear for very large N.}
\end{figure}

As in many ABM, the total number of agents is an important parameter on which the simulations output usually depend. Thanks to the model simplicity, we are able to understand the dynamics for different values of $N$. \\From eq.(\ref{eq:9}) we can see that for large $N$ fundamentalists dominate. Alfarano et al. \cite{AL2} have found that in the case $\delta=0$ eq.(\ref{eq:8}) leads to an alternation between the classes whose characteristic times grows linearly with $N$. \textbf{These two results make clear that only an intermediate value $N^*$ of $N$ can produce the SF}: in fact, if $N$ is too low the fluctuations between fundamentalists and chartists are too quick while if $N$ is too high the system gets stuck in the fundamentalist state. \\Figure \ref{fig:2} shows the three cases $N\ll N^*$, $N\gg N^*$ and $N\approx N^*$. Obviously, the specific $N^*$ depends on the other parameters of the model. Given the parameters used in \cite{paperoI,paperoEPL,paperoSOI}, $N^*$ is approximately $500$.  Therefore, a question spontaneously arises: why real financial markets go and persist in the state $N\approx N^*$? This is the self-organization problem, which we are going to address in the following section.
\subsection{\textbf{Self-Organized Criticality}}
\begin{figure}[ht]
\includegraphics[clip=true,keepaspectratio,width=12cm]{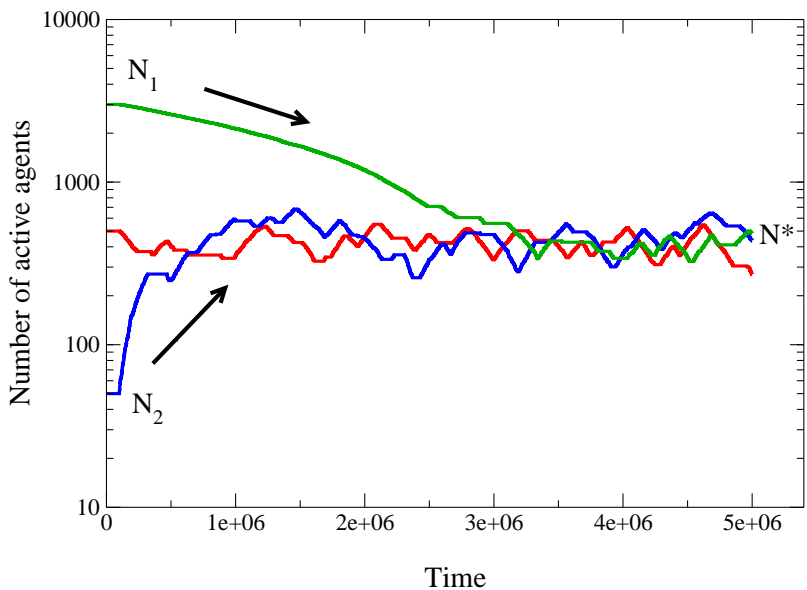}
\caption{\label{fig:5} 
Example of the self-organization of the system to the quasi-critical state
corresponding to the 
intermittent behavior and the SF.
Starting with a large number of agents ($N=5000$) the system is rather
stable in the fundamentalist state 
and price fluctuations are very small. This will induce agents to become
inactive and $N$ will consequently decrease (green line). If we start from a small number ($N=50$) fluctuations
are large and this 
will attract agents to enter this market (blue line). The critical number of
agents ($N=500$ for our choice of parameters) remains essentially stable (red line).}
\end{figure}

In Statistical Physics the presence of power laws is frequently associated with Critical Phenomena: in the thermodynamic limit $N\rightarrow \infty$ and thanks to a fine tuning of some parameters (typically the temperature) the system goes to a critical state in which power laws appear and exponents have well-defined values. Regarding financial markets, the situation is very different, because $N$ is a highly fluctuating variable and no explicit exogenous tuning exists, so that \textbf{the system must be spontaneously driven from a variety of initial conditions towards a critical state} (this phenomenon is called Self-Organized Criticality, SOC hereafter).\\
\textbf{ We argue that the key parameter to understand the SOC of financial markets is $N$}. In fact, as we have seen, only if $N$ is close to $N^*$ SF appear. We would like to stress that $N^*$ eventually depends on the specific features of the considered market, but it is, in any case, finite, and consequently one should talk about a \textit{quasi}-critical state, in which apparent power laws appear because of \textit{finite size effects}. In this context \textit{no universality must be expected}: this explains the different exponents observed for different markets \cite{bouchaud}. In order to investigate how and why $N \rightarrow N^*$ in real financial markets, we propose the following mechanism to introduce the temporal evolution of $N$.\\\\
Let us suppose $N\ll N^*$. This is a highly fluctuating market, in which however a lot of profitable opportunities appear: both fundamentalists and chartists see price signals, so agents will be encouraged to enter. On the contrary, if $N\gg N^*$ the market is fully efficient and this superstable state does not permit to gain significant profit from agents' strategies, and therefore agents will tend to exit from the market. In order to quantify and include this line of reasoning in the model, the long term volatility is considered:
\begin{equation}
\sigma(t, T)=\frac{1}{T-1}\sum_{i=t}^{t-T}(p_i-\bar p)^2
\label{eq:12}
\end{equation} 
where $T$ is very large with respect to the population dynamics time scale. An agent will enter the market if $\sigma(t, T)>\Theta_{in}$
and exit if $\sigma(t, T)<\Theta_{out}$, where $\Theta_{in}$ and $\Theta_{out}$ are suitable thresholds. 
In fig.\ref{fig:5} we show the temporal evolution of $N(t)$. The systems clearly self-organizes in the critical state in which $N\approx N^*$ and SF emerge spontaneously.  \\
The proposed mechanism could be apparently problematic because it may be argued that speculators could be scared by a too fluctuating market. 
Besides that this problem can be easily clarified by the analysis of fluctuations at different timescales. 
In \cite{alfisoi}, the authors investigate the robustness of this mechanism with respect to a change of thresholds and taking into account risk adverse agents which want to avoid large volatility at short timescale which would induce a high risk on strategies. \\
This line of reasoning could appear problematic if one believes that risk adverse agents could find a stable market more attractive than a turbulent one; however, this kind of agents are not the ones leading to the SF and are not explicitly considered in our model.

\subsection{\textbf{Realism and tractability}}
This model aims at finding a good compromise between realism and tractability. In fact, it maintains the essential elements of the very realistic LM model as a starting point and then introduces a drastic simplification of the price formation mechanism, chartists' strategies and opinion changes. This makes the dynamics analytically manageable and makes the behavior that this class of models shows when $N$ varies finally clear.\textbf{ This has opened the possibility to study the self-organization of financial markets and permitted the interpretation of the SF as finite size effects}.
\subsection{\textbf{Perspectives}}
The model simplicity can be easily exploited introducing more realistic features in a systematic way. The possible extensions include price reaction to an abrupt change in the fundamental price, different strategies, herding in a network, finite wealth and so on. Actually, the authors of the present review are studying the portfolio coherence emerging from collective lack of trust and its implications in terms of systemic risk. 
\section{Conclusions and perspectives}
We have discussed in an unifying framework some representative agent-based models for Economics and price dynamics. In particular we focus on the essential active principle of each one and discuss them in a systematic organization.
Broadly speaking one can define two variables: realism and tractability and in fig. \ref{revstr} we have given a pictorial representation of the models in this respect.\\
\subsection*{\textbf{Which model is better?}}
This question has no clear answer because the Stylized Facts, which represent the experimental framework are relatively limited and not too difficult to reproduce in ABM. Therefore a first conclusion is that progress in this area is directly related to the possibility of increasing the number and the quality of the Stylized Facts. This would provide additional tests of the models and permit a quality assessment of them.
\\
\subsection*{\textbf{How many parameters?}}
This question relates to the fundamental dilemma of realism versus simplicity and has no simple answer. Clearly in terms of metaphors the minimum number of parameters is optimal to grasp a concept in a simple way. This has been the philosophy of the fundamental models for critical phenomena in Statistical Physics and Complex Science \cite{baffoEN}. On the other hand, even a simplified description of real markets may require more elements specific to this field which may not have a counterpart in Statistical Physics.
\\
\subsection*{\textbf{Universality}}
The Stylized Facts observed, characterized essentially by fat tails and volatility clustering resemble the power law behavior of Critical Phenomena. This leads to the fundamental issue of Universality. Clean power laws in fact are observed as asymptotic properties and, in this limit, they may be universal in the sense that do not depend on the details of the model but only on its basic principles. Many physicists like the idea that the same may happen in Economics and that the observed deviations from universality in the data depend only on the poor quality of these.\\
In our opinion we are in an intermediate situation. Certain properties of the various markets appear qualitatively similar but their quantitative properties are really different. This also touches the question whether the power laws are approximate or genuine and clearly we point to the first hypothesis.\\
The basic reason for this point of view is that in most of the models considered the Stylized Facts do not correspond to a genuine asymptotic behavior and they are can be obtained for a very specific number of agents.\\
This situation of similarity but not strict universality has advantages and disadvantages. On one hand one looses the clean properties of universal exponents. On the other hand one acquires a sort of biodiversity which may lead to a specific characterization of the properties of each market.
\\
\subsection*{\textbf{How many agents and the problem of Self-Organization}}
In some agent-based models the number of effective agents is fixed at some finite value while in others it can vary in a more or less important way. In all models, however, the Stylized Facts are reproduced only in a very limited region of parameters. We believe that the issue of the self-organization is a crucial point which should be considered as an essential part of the model. Namely the dynamics of the system should be such that it leads spontaneously to the realistic region. We believe that the fluctuations in the effective number of agents or strategies are an important concept for the self-organization.

\acknowledgments
This work is partly supported by the scientific project FOC (Forecasting Financial Crises) financed by FET OPEN Scheme.

 \bibliographystyle{varenna}
 \bibliography{sample}

\begin{thebibliography}{10}
\expandafter\ifx\csname url\endcsname\relax\def\url#1{\texttt{#1}}\fi
\expandafter\ifx\csname urlprefix\endcsname\relax\def\urlprefix{URL }\fi

\bibitem{buchanan}
\NAME{Buchanan M.}, \IN{New Scientist}{2665}{2008}{32}.

\bibitem{kreps}
\NAME{Kreps D.~M.}, \TITLE{A course in microeconomic theory} (Princeton
  University Press) 1990.

\bibitem{bachelier}
\NAME{Bachelier L.}, \IN{Annales Scientifiques de l'Ecole Normale
  Sup\'erieureI}{II-17}{1900}{21}, [Ph.D. thesis is mathematics].

\bibitem{feller}
\NAME{Feller W.}, \TITLE{An Introduction to Probabbility and Its Applications,
  Vol. 1} (J. Wiley \& Sons, New York) 1968.

\bibitem{blacknscholes}
\NAME{Black F. \atque Scholes M.}, \IN{J. Polit. Econ.}{81}{1973}{637}.

\bibitem{hull}
\NAME{Hull J.~C.}, \TITLE{Options, Features and Other Derivatives}
  (Prentice-Hall, Upper Saddle River NJ) 1997.

\bibitem{cutler}
\NAME{Cutler D.~M., Poterba J.~M. \atque Summers L.~H.}, \IN{The Journal of
  Portfolio Management}{15}{1989}{4}.

\bibitem{baffoEN}
\NAME{Pietronero L.}, \IN{Europhysics News}{39}{2008}{26}.

\bibitem{bouchaudnature}
\NAME{Bouchaud J.-P.}, \IN{Nature}{455}{2008}{1181}.

\bibitem{farmernature}
\NAME{Farmer J.~D. \atque Foley D.}, \IN{Nature}{460}{2009}{685}.

\bibitem{rama}
\NAME{Cont R.}, \IN{Quant. Fin.}{1}{2001}{223}.

\bibitem{libroandrea}
\NAME{Harris L.}, \TITLE{Trading and exchanges} (Oxford University Press) 2003.

\bibitem{taka2}
\NAME{{Mizuno} T., {Takayasu} H. \atque {Takayasu} M.}, \IN{Physica
  A}{382}{2007}{187}.

\bibitem{taka1}
\NAME{{Takayasu} M., {Mizuno} T. \atque {Takayasu} H.}, \IN{Physica
  A}{370}{2006}{91}.

\bibitem{bouchaud}
\NAME{Bouchaud J.-P. \atque Potters M.}, \TITLE{Theory of Financial Risk and
  Derivative Pricing: From Statistical Physics to Risk Management} (Cambridge
  University Press) 2003.

\bibitem{benoit1}
\NAME{Mandelbrot B.}, \IN{Journal of Business}{35}{1963}{394}.

\bibitem{mantegnabook}
\NAME{Mantegna R.~N. \atque Stanley H.}, \TITLE{An Introduction to
  Econophysics: Correlation and Complexity in Finance} (Cambridge University
  Press, New York, NY, USA) 2000.

\bibitem{reviu_abergel}
\NAME{Chakraborti A., Toke I.~M., Patriarca M. \atque Abergel F.}, \IN{e-Print
  arXiv:0909.1974}{}{2009}{}.

\bibitem{lebraon_stat}
\NAME{LeBaron B.}, \IN{Quant. Fin.}{1}{2001}{621}.

\bibitem{rama_ref1}
\NAME{Andersen T.~G. \atque Bollerslev T.}, \IN{J. Empirical
  Finance}{4}{1998}{115}.

\bibitem{rama_ref2}
\NAME{An\'e T. \atque Geman H.}, \IN{J. Risk}{2}{1999}{57}.

\bibitem{rama_ref3}
\NAME{Bak P., Tang C. \atque Wiesenfeld K.}, \IN{Phys. Rev.
  Lett.}{59}{1987}{4}.

\bibitem{rama_ref4}
\NAME{Mizuno T., Takayasu M. \atque Takayasu H.},
  \IN{http://arxiv.org/pdf/physics/0508162}{}{2005}{}.

\bibitem{arrow}
\NAME{Arrow J.~K. \atque Debreu G.}, \IN{Econometrica}{22}{1954}{265}.

\bibitem{ref_far1}
\NAME{Debreu G.}, \TITLE{Theory of Value.} (Wiley NY) 1959.

\bibitem{ref_far2}
\NAME{Arrow J.~K.}, \IN{Econometrie}{11}{1953}{41}.

\bibitem{sf2}
\NAME{Arthur W., Holland J.~H., LeBaron B., Palmer R. \atque Tyler P.},
  \IN{http://www2.econ.iastate.edu/tesfatsi/ahlpt96.pdf}{}{1996}{}.

\bibitem{reviu}
\NAME{Samanidou E., Zschischang E., Stauffer D. \atque Lux T.}, \IN{Reports on
  Progress in Physics}{70}{2007}{}.

\bibitem{reviu_hommes}
\NAME{Hommes C.~H.}, Tinbergen Institute Discussion Papers Tinbergen Institute
  05-056/1, Tinbergen Institute (2005).
\newline\urlprefix\url{http://ideas.repec.org/p/dgr/uvatin/20050056.html}

\bibitem{black2}
Tech. Rep. The Brady Commission US Government Commission Printing Office, The
  Brady Commission, Washington DC (1988).

\bibitem{black1}
\NAME{Schwert G.~W.}, \IN{Review of Financial Studies}{3}{1990}{77}.

\bibitem{flashcrash}
Tech. Rep. SEC, SEC (2010).
\newline\urlprefix\url{http://www.sec.gov/sec-cftc-prelimreport.pdf}

\bibitem{rational}
\NAME{Diba B.~T. \atque Grossman H.~I.}, \IN{The Economic
  Journal}{98}{1988}{746}.

\bibitem{KIM}
\NAME{Kim G. \atque Markowitz H.~M.}, \IN{Journal of Portfolio
  Management}{16}{1989}{45}.

\bibitem{portfolio_en}
\NAME{Brennan M.~J. \atque Schwartz E.~S.}, \IN{Journal of
  Business}{62}{1989}{455}.

\bibitem{sf1}
\NAME{Palmer R., Arthur W., Holland J.~H., LeBaron B.,  \atque Tyler P.},
  \IN{Physica D}{75}{1994}{264}.

\bibitem{libropatterns}
\NAME{Roehner B.~M.}, \TITLE{Patterns of Speculation} (Cambridge Univeristy
  Press) 2002.

\bibitem{Challet:2005:MGI:1121663}
\NAME{Challet D., Marsili M. \atque Zhang Y.-C.}, \TITLE{Minority Games:
  Interacting Agents in Financial Markets} (Oxford University Press, Inc., New
  York, NY, USA) 2005.

\bibitem{Andrea2006:0305-4470:R465}
\NAME{{De Martino} A.}, \IN{Journal of Physics A: Mathematical and
  General}{39}{2006}{R465}.

\bibitem{bar}
\NAME{Arthur B.~W.}, \IN{Am. Econ. Rev.}{84}{1994}{406}.

\bibitem{mingame}
\NAME{Challet D. \atque Zhang Y.}, \IN{Physica A}{246}{1997}{407}.

\bibitem{minstil}
\NAME{Challet D., Marsili M. \atque Zhang Y.-C.}, \IN{Physica A: Statistical
  Mechanics and its Applications}{}{2001}{514}.

\bibitem{savit}
\NAME{{Li} Y., {VanDeemen} A. \atque {Savit} R.}, \IN{Physica A Statistical
  Mechanics and its Applications}{284}{2000}{461}.

\bibitem{PhysRevE.68.036132}
\NAME{Challet D. \atque Marsili M.}, \IN{Phys. Rev. E}{68}{2003}{036132}.

\bibitem{challet:marsili:zecchina}
\NAME{Challet D., Marsili M. \atque Zecchina R.}, \IN{Phys. Rev.
  Lett.}{84}{2000}{1824}.

\bibitem{citeulike:1398137}
\NAME{Mezard M., Parisi G. \atque Virasoro M.} 1986.

\bibitem{caldase}
\NAME{Caldarelli G., Marsili M. \atque Zhang Y.-C.}, \IN{Eurphys.
  Lett.}{40}{1997}{479}.

\bibitem{calda_critica}
\NAME{Golke S., Preis T., Paul W. \atque Schneider J.~J.},
  \IN{http://www.staff.uni-mainz.de/schneidj/papers/papergolke.pdf}{}{}{}.

\bibitem{LMnature}
\NAME{Lux T. \atque Marchesi M.}, \IN{Nature}{397}{1999}{498}.

\bibitem{LM}
\NAME{Lux T. \atque Marchesi M.}, \IN{International Journal of Theoretical and
  Applied Finance}{3}{2000}{675}.

\bibitem{kirman}
\NAME{Kirman A.}, \IN{The Quarterly Journal of Economics}{108}{1993}{137}.

\bibitem{alfasi}
\NAME{Alfarano S., Lux T. \atque Wagner F.}, \IN{Comput. Econ.}{26}{2005}{19}.

\bibitem{paperoI}
\NAME{Alfi V., Cristelli M., Pietronero L. \atque Zaccaria A.}, \IN{Eur. Phys.
  J. B}{67}{2009}{385}.

\bibitem{Zawadowski2002403}
\NAME{Zawadowski A.~G., Karadi R. \atque Kertesz J.}, \IN{Physica A:
  Statistical Mechanics and its Applications}{316}{2002}{403 }.

\bibitem{LMfinite}
\NAME{Egenter E., Lux T. \atque Stauffer D.}, \IN{Physica A}{268}{1999}{250}.

\bibitem{paperoEPL}
\NAME{Alfi V., Pietronero L. \atque Zaccaria A.}, \IN{EPL}{86}{2009}{58003}.

\bibitem{paperoII}
\NAME{Alfi V., Cristelli M., Pietronero L. \atque Zaccaria A.}, \IN{Eur. Phys.
  J. B}{67}{2009}{399}.

\bibitem{giard_bou}
\NAME{Giardina I. \atque Bouchaud J.-P.}, \IN{Eur. Phys. J. B}{31}{2003}{421}.

\bibitem{lillocoll}
\NAME{Lillo F., Farmer J.~D. \atque Mantegna R.~N.},
  \IN{Nature}{421}{2003}{129}.

\bibitem{revBOU}
\NAME{Bouchaud J.-P., Farmer J.~D. \atque Lillo F.}, \TITLE{How Markets Slowly
  Digest Changes in Supply and Demand.} (Elsevier: Academic Press) 2008.

\bibitem{giard2}
\NAME{Bouchaud J.-P., Giardina I. \atque M\'ezard M.}, \IN{Quant.
  Fin.}{1}{2001}{212}.

\bibitem{paperoSOI}
\NAME{Alfi V., Cristelli M., Pietronero L. \atque Zaccaria A.}, \IN{J. Stat.
  Mech.}{}{2009}{P03016}.

\bibitem{thurner}
\NAME{Thurner S., Farmer J.~D. \atque Geanakoplos J.}, \IN{Cowles Foundation
  Discussion Paper}{1745}{2010}{}.
\newline\urlprefix\url{http://dido.econ.yale.edu/P/cd/d17a/d1745.pdf}

\bibitem{RePEc:spr:jeicoo:v:4:y:2009:i:2:p:195-212}
\NAME{{Delli Gatti} D., Gallegati M., Greenwald B., Russo A. \atque Stiglitz
  J.}, \IN{Journal of Economic Interaction and Coordination}{4}{2009}{195}.

\bibitem{alfi2}
\NAME{Alfi V., Coccetti F., Marotta M., Pietronero L. \atque Takayasu M.},
  \IN{Physica A}{370}{2006}{30}.

\bibitem{alfi3}
\NAME{Alfi V., {De Martino} A., Pietronero L. \atque Tedeschi A.}, \IN{Physica
  A}{382}{2007}{1}.

\bibitem{AL2}
\NAME{S.Alfarano, T.Lux \atque F.Wagner}, \IN{Journal of Economic Dynamics {\&}
  Control}{32}{2008}{101}.

\bibitem{alfisoi}
\NAME{Alfi V., Cristelli M., Pietronero L. \atque Zaccaria A.}, \IN{J. Stat.
  Mech.}{}{2009}{P03016}.

\end{thebibliography}

\end{document}